\documentclass[12pt]{article}
\usepackage{amsfonts,graphicx,amssymb,amsmath,amsthm,epsfig}
\usepackage{float}
\allowdisplaybreaks
\begin{document}
\title{\textbf{Exploring Hybrid Star Models with Quark and Hadronic Matter in
$f(Q)$ Gravity}}
\author{M. Sharif$^{1,2}$\thanks{msharif.math@pu.edu.pk (Corresponding author)}~~and
Madiha Ajmal$^1$\thanks{madihaajmal222@gmail.com} \\
$^1$ Department of Mathematics and Statistics, The University of Lahore,\\
1-KM Defence Road Lahore-54000, Pakistan.\\
$^2$ Research Center of Astrophysics and Cosmology, Khazar
University,\\ Baku, AZ1096, 41 Mehseti Street, Azerbaijan.}

\date{}
\maketitle
\begin{abstract}
In this paper, we develop a model for a static anisotropic hybrid
star that includes strange quark matter and hadronic matter. We
solve the field equations in the $f(Q)$ gravity framework (where $Q$
is the non-metricity) using the Finch-Skea metric.  The relationship
between density and pressure for strange quark matter is described
using the MIT bag model equation, while for hadronic matter, the
radial pressure and density are related by a linear equation of
state. We select the compact star EXO 1785-248 and analyze five
different values of the coupling constant. To evaluate the physical
feasibility of the model, we perform a graphical analysis of key
properties, including the metric components, energy density, radial
and tangential pressures, anisotropy, gradients, quark matter
density and pressure, the equation of state parameter, energy
conditions and the mass function. We further examine the stability
and equilibrium of the star through parameters such as compactness,
redshift, causality conditions, Herrera cracking, the adiabatic
index and the Tolman-Oppenheimer-Volkoff equation. We observe that
$f(Q)$ gravity effectively describes the macroscopic properties of
hybrid stars.
\end{abstract}

\section{Introduction}

General Relativity (GR) has been a powerful and reliable framework
for understanding gravity. However, observations of the rapid
expansion of the universe have revealed certain limitations of GR.
In addition to the universe's accelerated expansion, several other
observational and theoretical issues also highlight the limitations
of GR. For example, the rotation curves of galaxies cannot be
explained by GR without assuming the existence of dark matter.
Similarly, anomalies in gravitational lensing, cosmic microwave
background data and large scale structure formation challenge the
completeness of GR. Moreover, GR breaks down near singularities such
as in black holes and is not renormalizable in quantum field theory.
These issues collectively motivate the exploration of modified
gravity theories (MGTs) as potential extensions or alternatives to
GR \cite{1}. The accelerated expansion is a pivotal aspect of the
history of the universe, prompting questions about the mysterious
force responsible for it. Since GR alone does not fully explain this
phenomenon, scientists are exploring new theories to solve this
mystery \cite{2}.

There are two main approaches to explain the universe accelerated
expansion. The first approach suggests modifying the universe
composition by introducing new elements, like dark energy (DE). Dark
energy is characterized by high negative pressure and remains
undetected in direct observations. Einstein introduced the
cosmological constant ($\Lambda$) in his equations as a leading
candidate for DE. It aligns well with current observations, though
it also presents some theoretical challenges. The second approach is
to modify GR itself through MGTs. Recent experiments and
observations indicate that some MGTs effectively describe both the
early universe inflation and later-period acceleration. With these
two approaches, scientists aim to understand the accelerated
expansion and possibly unveil hidden aspects of the universe.

In GR, gravity is explained through the Ricci scalar ($R$), which
arises from Riemannian geometry based on the curvature of spacetime.
An extension of GR, known as $f(R)$ gravity, generalizes the theory
by replacing $R$ with an arbitrary function of $R$. The GR
traditionally relies on the curvature of the Levi-Civita connection
\cite{5}, which is free from torsion. Other gravitational theories,
such as the teleparallel equivalent of GR (TEGR) \cite{6}, replaces
curvature with torsion $\mathcal{T}$, leading to a modified version
known as $f(\mathcal{T})$ gravity. Yet another approach is symmetric
teleparallel gravity (STG) \cite{7}, which describes gravity via the
non-metricity, eliminating both curvature and torsion. Extending
this framework, $f(Q)$ gravity \cite{8} replaces $Q$ with a general
function of $Q$, opening new possibilities for understanding
gravitational phenomena, especially in cosmology and astrophysics.
This theory has attracted attention to explain cosmic acceleration
and other large-scale cosmic dynamics \cite{9}.

From an astrophysical and cosmological point of view, $f(Q)$ gravity
offers several distinct advantages over more commonly studied MGTs
such as $f(R)$ and $f(\mathcal{T})$ gravity theories. These theories
offer innovative frameworks for understanding the complex structures
of hybrid stars, which consist of a mixture of strange quark matter
and hadronic matter. While $f(R)$ and $f(\mathcal{T})$ gravity
theories have been extensively explored to address various cosmic
phenomena, $f(Q)$ gravity introduces key advantages, particularly in
modeling extreme stellar environments and handling anisotropic
pressures. The $f(R)$ theory has been successful in addressing
large-scale cosmic issues such as DE and the expansion of the
universe \cite{66b}, it faces challenges when it comes to modeling
singularities and exotic matter states, particularly in the context
of stellar interiors \cite{66c}. Similarly, $f(\mathcal{T})$ gravity
\cite{66d}, which is based on torsion instead of curvature, offers a
distinct perspective on gravitational dynamics \cite{66e} but
struggles to handle the complexity of hybrid star interiors in the
same way $f(Q)$ does.

One widely studied theory is $f(Q)$ gravity, which has shown broad
potential in making predictions about both astrophysical and
cosmological phenomena. Anagnostopoulos et al. \cite{13} introduced
a new $f(Q)$ model that, while similar to $\Lambda$CDM in
parameters, offers unique cosmological behavior. Frusciante
\cite{14} examined a model within this theory that resembles the
large-scale behavior of the $\Lambda$CDM model but shows unique
characteristics at the level of linear perturbations. Mandal et al.
\cite{16} studied how bulk viscosity impacts cosmological models.
Wang et al. \cite{18} studied spherically symmetric fluid spheres in
this theory, finding that a Schwarzschild anti-de Sitter solution
exists but an exact Schwarzschild solution is not possible for
nontrivial $f(Q)$ functions. D'Ambrosio et al. \cite{19} developed
perturbative corrections for Schwarzschild solutions in this
gravity, incorporating extra characteristics arising from the
dynamic affine connection. Gadbail et al. \cite{20} proposed various
$f(Q)$ gravity models, demonstrating that $Q$ can replicate the
observed accelerated cosmic expansion similar to $\Lambda$CDM.
Recent papers \cite{21} have extensively examined $f(Q)$ gravity,
investigating various geometrical and physical aspects of this
theory.

A hybrid star is a unique astrophysical object that combines the
properties of neutron stars and quark stars, representing a
transitional phase between the two. These stars emerge when a
massive star undergoes gravitational collapse after exhausting its
nuclear fuel, leading to the formation of a dense neutron rich core.
In the core of a hybrid star, the density becomes so extreme that
neutrons are decomposed into their constituent quarks, resulting in
a state known as deconfined quark matter. This quark matter often
includes strange quarks, further emphasizing the hybrid nature of
the star. The outer layers of a hybrid star remain similar to those
of a traditional neutron star, composed primarily of nuclear matter
such as neutrons and protons bound by strong nuclear forces. This
unique composition makes hybrid stars essential for studying matter
under extreme conditions, compact objects and high energy
astrophysical phenomena. Among the models, the conventional hybrid
star featuring quark matter at the core and neutron matter in the
crust is preferred due to its alignment with observational data,
including GW170817 tidal deformability constraints, NICER radius
measurements and massive pulsar masses. In contrast, the inverted
hybrid model \cite{21a}, with hadronic matter at the core and quark
matter in the crust, relies on speculative assumptions such as quark
matter stability at low pressures, which lacks observational
confirmation. Thus, the conventional model stands out for its
reliability, extensive study and compatibility with astrophysical
constraints.

Mishra et al. \cite{22} analyzed neutron-quark phase transitions in
hybrid quark stars, finding stable, predominantly quark matter stars
with a neutron crust by solving Tolman-Oppenheimer-Volkoff
equations. Khadkikar et al. \cite{24} studied neutron-to-quark
matter phase transitions in hybrid stars using a relativistic
confinement model, achieving stable solutions. Maheswari et al.
\cite{25} proposed an equation of state (EoS) incorporating
spin-isospin forces, aligning well with unpolarized nuclear matter.
This EoS predicts a ferromagnetic transition in neutron matter at
high densities and effectively models the mass, size and magnetic
fields of hybrid stars. Schertler et al. \cite{26} examined medium
effects in quark matter, finding that they reduced the pure quark
matter core in hybrid stars, favoring a mixed quark-hadronic phase
across various strong coupling constants. Blackman et al. \cite{28}
explored the effect of star-disk magnetohydrodynamic winds on
planetary nebulae, proposing that disk-dominated winds and magnetic
shaping explain multi-polar structures. Gupta et al. \cite{29}
studied the impact of the quark-nuclear matter mixed phase on radial
oscillation modes in hybrid stars using relativistic mean-field
theory and realistic nucleon interactions. Grigorian et al.
\cite{30} derived EoS for quark matter using a nonlocal chiral quark
model, emphasizing diquark condensation's impact on color
superconductivity and hybrid star stability.

Alford et al. \cite{30a} explored how hybrid stars resemble the
mass-radius relationship of purely nucleonic stars using realistic
EoS. Nicotra et al. \cite{30b} investigated the hadron-quark phase
transition in protoneutron stars, finding maximum masses under 1.6
solar masses. Hussain et al. \cite{31} synthesized hybrid
fluorinated star polymers, resulting in enhanced thermal stability,
spherical particle formation and improved surface properties.
Dexheimer et al. \cite{34} demonstrated that high magnetic fields
alter the particle composition, structure and anisotropy of hybrid
stars. Alford et al. \cite{34a} studied hybrid star mass-radius
curves, finding conditions where hybrid stars exceed 2 solar masses.
Bhar \cite{35} proposed a new hybrid star model that matches
observational mass data and requires negative surface pressure for
stability. Burgio and Zappala \cite{36} analyzed the color-flavor
locking phase in hybrid star quark cores, noting its phase
transition features and the influence of hyperons on mass-radius
relations. Kaltenborn et al. \cite{36a} proposed a hybrid EoS with
excluded volume effects, predicting third-family compact stars
meeting the 2M$\odot$ constraint. Nandi and Char \cite{37} found
that the GW170817 tidal deformability constraint limits MIT bag
model parameters and sets upper radius limits for 1.4 and 1.6
$\mathcal{M}_\odot$ hybrid stars. Khanmohamadi et al. \cite{39}
found that hybrid star models with hadron-quark phase transitions
satisfy GW170817 constraints on radii and tidal deformability,
achieving stable high-mass configurations. Recent studies have made
significant advancements in understanding hybrid stars \cite{42}.

The study of hybrid stars in MGT is crucial due to their unique
environments, which provide insights into extreme astrophysical
phenomena. These stars, composed of quark and hadronic matter, exist
at densities exceeding nuclear levels, offering a natural laboratory
to explore high-density matter and the hadronic-to-quark phase
transition. MGTs, which extend GR to address challenges like DE and
the universe's accelerated expansion, enable researchers to
investigate gravity in strong-field regimes. In this context, MGT
predicts solutions that avoid singularities, proposing smooth core
structures for hybrid stars, which differ from the predictions of
GR. Observational data on hybrid star's masses, radii and structures
further refine our understanding of gravity in extreme conditions,
bridging weak and strong gravitational fields. Abbas and Nazar
\cite{66b} studied stable, singularity-free anisotropic compact
stars with quark and hadronic matter under $f(R)$ gravity, using the
bag model EoS. Bhar et al. \cite{40} investigated quark
deconfinement in hybrid star cores, modeling phase transitions and
analyzing the impact of bag pressure on star structure, with
findings compared to observational data. Rej \cite{40a} examined a
hybrid stellar model of quark and hadronic matter in modified $f(G)$
gravity ($G$ denotes the Gauss-Bonnet invariant term) and evaluated
its physical validity for the compact star 4U 1538-52.

The Finch-Skea metric has an important role for modeling compact
stars in four or higher dimensional spacetime. Its non-singular and
practical properties make it widely adopted in compact star
modeling, leading to its frequent applications in compact stellar
objects with various forms of matter. This metric has been
extensively used to examine gravastars, wormholes, neutron stars,
hybrid stars and strange stars in both GR and alternative gravity
theories. Finch and Skea \cite{43} developed a new mathematical
model (metric) for describing spherical, symmetric, compact stars.
Later, this Finch-Skea metric was adapted to four dimensions to
create models of stars with uneven (anisotropic) pressures
\cite{44}. Sharma and Ratanpal \cite{45} used the Finch-Skea metric
to propose a model for relativistic stars. Sharma and Das \cite{46}
explored Bessel function-based solutions for anisotropic systems
within the Finch-Skea spacetime. Pandya et al. \cite{47} expanded
the Finch-Skea framework to propose new solutions for spherically
symmetric anisotropic matter, showing compatibility with various
compact stars. Molina et al. \cite{50} demonstrated that static
fluid solutions in pure Lovelock gravity exhibit universal behavior
across dimensions, unifying stellar interior models from the compact
Schwarzschild to Finch-Skea states. Banerjee et al. \cite{52}
examined how DE affects compact astrophysical objects using the
Finch and Skea ansatz to solve Einstein equations.

Dayanandan et al. \cite{f2bb} explored a modified Finch-Skea
anisotropic configuration within Class I spacetime by employing
gravitational decoupling to construct viable models of compact
stars. Gul et al. \cite{f2aa} assessed the physical acceptability
and stability of anisotropic stellar models in $f(Q)$ gravity using
Finch-Skea geometry, validating the models through several stability
checks. Mustafa et al. \cite{3gg} analyzed the behavior of a static,
spherically symmetric anisotropic star under the framework of $f(Q)$
theory by incorporating the Karmarkar condition along with the
Finch-Skea structure. Rej et al. \cite{f61} employed the complexity
factor approach with a Finch-Skea background to develop a
relativistic model of an anisotropic DE star. Shahzad et al.
\cite{f62} proposed a new solution in Rastall gravity coupled with a
quintessence field, applying the Finch-Skea framework to study
isotropic compact stars. Das et al. \cite{f64} examined anisotropic
relativistic configurations with the Finch-Skea metric, particularly
focusing on applications to the pulsar PSR J0348+0432.

In this study, we focus on developing a hybrid star model within the
framework of $f(Q)$ gravity to incorporate recent observations of
compact stars. The paper is organized as follows. Section \textbf{2}
provides an overview of $f(Q)$ gravity and also discusses the
internal spacetime. Section \textbf{3} solves the field equations
using the Finch and Skea ansatz metric. Section \textbf{4} explains
the connection between the inner spacetime and the outer
Schwarzschild vacuum solution at the boundary $r=\mathcal{R}$ and
expresses the metric constants in terms of the star's mass and
radius. In sections \textbf{5} and \textbf{6}, we discuss some
physical properties and stability of the model, respectively.
Section \textbf{7} presents the summary of the results obtained.

\section{Formulation of $f(Q)$ Gravity}

The affine connection $\hat{\Gamma}^{\mu}_{\nu\varsigma}$ in the
framework of $f(Q)$ gravity can be expressed as a combination of
three independent components \cite{53a}
\begin{equation}\label{s1}
\hat{\Gamma}^{\mu}_{\nu\varsigma}={\Gamma}^{\mu}_{\nu\varsigma}
+K^{\mu}_{\nu\varsigma}+\mathbb{L}^{\mu}_{\;\nu\varsigma}.
\end{equation}
where $\Gamma^{\mu}_{\nu\varsigma}$ is the Levi-Civita connection,
$K^{\mu}_{\nu\varsigma}$ is the contortion tensor and
$\mathbb{L}^{\mu}_{\nu\varsigma}$ is the disformation tensor. The
contortion tensor is defined as
\begin{equation}\label{s2}
K^\mu_{\nu\varsigma} = \hat{\Gamma}^{\mu}_{[\nu\varsigma]} +
g^{\mu\sigma}g_{\nu\kappa}\hat{\Gamma}^{\kappa}_{[\varsigma\sigma]}
+g^{\mu\sigma}g_{\varsigma\kappa}\hat{\Gamma}^{\kappa}_{[\nu\sigma]},
\end{equation}
while the disformation tensor is given by
\begin{equation}\label{s3}
\mathbb{L}^{\mu}_{\nu\varsigma} =
\frac{1}{2}g^{\mu\sigma}(Q_{\varsigma\nu\sigma} +
Q_{\nu\varsigma\sigma} - Q_{\mu\nu\varsigma}),
\end{equation}
with the non-metricity tensor defined as $Q_{\varsigma\nu\sigma} =
\nabla_{\sigma}g_{\nu\varsigma}$. The Levi-Civita connection can be
written as
\begin{equation}\label{s4}
\Gamma^{\mu}_{\nu\varsigma} =
\frac{1}{2}g^{\mu\sigma}(g_{\sigma\varsigma,\nu} +
g_{\sigma\nu,\varsigma} - g_{\nu\varsigma,\sigma}),
\end{equation}
and in the context of STG, it is related to the disformation tensor
by
\begin{equation}\label{s5}
\Gamma^{\mu}_{\nu\varsigma} = -\mathbb{L}^{\mu}_{\;\nu\varsigma}.
\end{equation}

The gravitational action in non-covariant form is expressed as
\begin{equation}\label{s6}
S=\frac{1}{2\kappa}\int g^{\nu\varsigma}(\Gamma^{\imath}_{\sigma\nu}
\Gamma^{\sigma}_{\varsigma\imath} -\Gamma^{\imath}_{\sigma\imath}
\Gamma^{\sigma}_{\nu\varsigma})\sqrt{-g} d^ {4}x,
\end{equation}
where $\kappa$ is the gravitational coupling constant, defined as
$\kappa = 8\pi G$, with $G$ being the gravitational constant. Using
Eq.\eqref{s5}, the action becomes
\begin{equation}\label{s7}
S=-\frac{1}{2\kappa} \int
g^{\nu\varsigma}(\mathbb{L}^{\imath}_{~\sigma\nu}
\mathbb{L}^{\sigma}_{~\varsigma\imath} -
\mathbb{L}^{\imath}_{~\sigma\imath}
\mathbb{L}^{\sigma}_{~\nu\varsigma}) \sqrt{-g} d^ {4}x.
\end{equation}
which corresponds to the action of STG. In this study, we adopt the
coincident gauge, in which the affine connection
$\hat{\Gamma}^{\mu}_{\nu\varsigma}$ vanishes identically. This gauge
choice is standard in STG and simplifies the formulation by
rendering the connection's equations of motion trivial. As a result,
only the metric degrees of freedom contribute dynamically, while the
connection does not introduce additional dynamics. This
simplification is justified for the current static, spherically
symmetric stellar model, and we clearly state that the effects of
connection dynamics are not considered in this work. The action for
$f(Q)$ gravity can be expressed as \cite{8}
\begin{equation}\label{1}
S=\int\bigg(\frac{1}{2\kappa}f(Q)+\mathcal{L}_{\mathbf{m}}\bigg)
\sqrt{-g}d^{4}x,
\end{equation}
where $g$ denotes the determinant of the metric tensor,
$\mathcal{L}_{\mathbf{m}}$ is the Lagrangian density for matter. The
field equations for $f(Q)$ gravity are then given by
\begin{equation}\label{6}
\frac{-2}{\sqrt{-g}}\nabla_{\mu}(f_{Q}\sqrt{-g}
P^{\mu}_{~\gamma\psi})-\frac{1}{2}f g_{\gamma\psi}-f_{Q}
(P_{\gamma\mu\nu}Q_{\psi}^{~\mu\nu}-2Q^{\mu\nu}_{~~~\gamma}
P_{\mu\nu\psi})=\kappa T_{\gamma\psi},
\end{equation}
where all relevant terms are detailed in Appendix \textbf{A} and
$f_{Q}=\frac{\partial f(Q)}{\partial Q}$.

We model the star interior spacetime using a static,
self-gravitating sphere defined by
\begin{equation}\label{7}
ds_-^{2}=e^{\alpha(r)}dt^{2}-e^{\beta(r)}dr^{2}-r^{2}(d\theta^{2}-\sin^{2}\theta
d\phi^{2}),
\end{equation}
where minus sign denotes the interior spacetime. This setup allows
us to describe the inner structure of a self-gravitating object in a
co-moving frame, where the fluid inside the star behaves as an
anisotropic relativistic fluid. The study considers a hybrid star
that contains two types of matter: hadronic matter and strange quark
matter. This model is based on the idea that under certain
conditions within a star, a phase transition might occur, where
normal nuclear matter changes into strange quark matter \cite{1ab}.
The energy-momentum tensor (EMT) associated with the two-fluid model
is given by the following notation
\begin{equation}\label{9}
T^0_0 = \rho + \rho_q, \quad T^1_1 = -(p_\mathbf{r} + p_\mathbf{q}),
\quad T^2_2 = T^3_3 = -(p_t + p_\mathbf{q}),
\end{equation}
where, $\rho$, $p_\mathbf{r}$ and $p_t$ represent the density,
radial pressure and tangential pressure, respectively, while
$\rho_\mathbf{q}$ and $p_\mathbf{q}$ denote the density and pressure
associated with quark matter. The field equations for a hybrid star
in $f(Q)$ gravity can be expressed as
\begin{eqnarray}\nonumber
\kappa(\rho + \rho_\mathbf{q}) &=& \frac{e^{-\beta}}{2r^2}\bigg[
f_{Q} \bigg((e^{\beta} - 1)(2 + r \alpha^{\prime}) + (1 + e^{\beta})
r \beta^{\prime} \bigg)+f r^2 e^{\beta}\\\label{10} &+& 2r f_{QQ}
Q^{\prime} (e^{\beta} - 1)\bigg],
\\\nonumber \kappa(p_\mathbf{r} + p_\mathbf{q}) &=& -  \frac{e^{-\beta}}{2r^2}\bigg
[f_{Q} \bigg((e^{\beta} - 1)(2 + r\beta^{\prime} + r
\alpha^{\prime}) - 2r \alpha^{\prime} \bigg)+f r^2
e^{\beta}\\\label{11} &+&2r f_{QQ} Q^{\prime} (e^{\beta} - 1)
\bigg],
\\\nonumber \kappa(p_t + p_\mathbf{q} )&=& \bigg
[f_\mathbf{q} \bigg(2\alpha^{\prime}(e^{\beta} - 2) - r
\alpha^{2\prime} + \beta^{\prime} (2e^{\beta} + r \alpha^{\prime}) -
2r \alpha^{\prime\prime} \bigg)+2f r e^{\beta}
\\\label{12} &-&2r f_{QQ} Q^{\prime} \alpha^{\prime}
\bigg]\frac{-e^{-\beta}}{4r} ,
\end{eqnarray}
where prime symbol denotes a derivative with respect to the radial
coordinate $r$. The scalar $Q$ becomes
\begin{equation}\label{14}
Q = \frac{1}{r}(e^{-\beta} - 1)(\alpha^{\prime} + \beta^{\prime}),
\end{equation}
where $\alpha^{\prime}$ and $\beta^{\prime}$ denote the derivatives
of the metric functions $\alpha(r)$ and $\beta(r)$, respectively,
with respect to the radial coordinate $r$.

The quadratic form of $f(Q)$ gravity has garnered significant
attention due to its potential to describe diverse phenomena in
cosmology and astrophysics. Mandal et al. \cite{aa} showed that the
$f(Q)$ gravity theory was validated by energy conditions, which
constrained $f(Q)$ models to those consistent with the universe
accelerated expansion. Khyllep et al. \cite{bb} revealed
$\Lambda$CDM-like dynamics in $f(Q)=Q +\alpha Q^n$ gravity, along
with deviations in matter growth and integrability. Lin and Zhai
\cite{cc} investigated the $f(Q)=Q+\alpha Q^2$ model, showing its
capability to support higher stellar masses with negative
modifications while reducing them with positive ones. Zhao \cite{dd}
explored the conflict between gauge choices and symmetry-based
coordinate systems, focusing on the quadratic model within this
framework. We adopt a quadratic form of $f(Q)$ gravity to describe
hybrid stars. This is expressed as
\begin{equation}\label{13}
f(Q)=Q+\zeta Q^{2},
\end{equation}
where $\zeta$ is a coupling constant that influences the gravity
model and its unit is $L^2$. We have chosen this form because it is
simple and mathematically manageable, effectively describing the
properties of compact stars such as hybrid stars. This form combines
linear and quadratic terms, providing deeper insight into
gravitational effects and the structure of such stars.

\section{Model of Hybrid Star}

We consider the Finch-Skea metric potentials which are given as
\cite{43}
\begin{equation}\label{15}
e^{\alpha(r)}=\bigg(\xi +\frac{1}{2} r \varphi  \sqrt{r^2 \chi
}\bigg)^2,\quad e^{\beta(r)}=r^2 \chi +1,
\end{equation}
where $\xi$, $\varphi$ and $\chi$ are unspecified, non-zero
constants. To complete the system, an additional constraint is
required, i.e., a well-founded relationship between $p_\mathbf{r}$
and $\rho$ of hadronic matter. There are multiple ways to express
this relationship. In our model, we adopt a linear EoS defined by
\begin{equation}\label{16}
p_\mathbf{r} = s\rho-h,
\end{equation}
where $0 < s < 1$ with $s \neq 1/3$, and $h > 0$. This EoS has been
used by various authors to model compact stars \cite{55}, providing
a solid basis for our approach. We also assume that the
pressure-density relationship for quark matter is described by the
MIT bag model EoS \cite{56}
\begin{equation}\label{17}
p_\mathbf{q} = \frac{1}{3}(\rho_\mathbf{q} - 4\mathcal{B}_g),
\end{equation}
where $\mathcal{B}_g$ represents the bag constant in units of
$MeV/fm^3$ \cite{57}. Witten \cite{58} suggested that strange quark
matter could be the true ground state of strongly interacting
matter. Farhi and Jaffe \cite{59} supported this idea, showing that
it holds for massless, non-interacting quarks when the bag constant
ranges between 57 and 94 $MeV/fm^3$. The main difference between
Eq.\eqref{16} and Eq.\eqref{17} lies in their physical context.
Eq.\eqref{16} is a general linear EoS used to model various types of
compact stars, with $s$ and $h$ providing flexibility to describe
different stellar matter. In contrast, Eq.\eqref{17} specifically
describes quark matter, derived from the MIT bag model, where the
bag constant $\mathcal{B}_g$ has a direct physical meaning related
to quantum chromodynamics (QCD) vacuum properties.

The linear EoS and MIT bag model \cite{59a} are widely used due to
their simplicity and reliability in modeling compact stars,
particularly quark-hadron phase transitions and hybrid stars. While
advanced models like the polytropic EoS, Nambu-Jona-Lasinio model
\cite{59b} and perturbative QCD \cite{59c} offer detailed
microscopic insights and nonlinear behavior, they are often more
complex and computationally demanding. Different models influence
mass-radius relationships, stability, and thermal evolution but the
MIT bag model's adjustable parameters and phenomenological approach
make it practical for studying observed phenomena such as tidal
deformability from GW170817. This balance between simplicity and
accuracy makes it a preferred choice for this study.

With these assumptions understood, we are now ready to solve the
field equations and derive expressions for the matter density and
pressure associated with regular hadronic matter as well as the
density and pressure associated with quark matter. The expressions
can be found in Appendix \textbf{B}.

\section{Boundary Conditions}

To ensure that spacetime remains continuous both inside and outside
the star model, it is essential for the inner and outer regions to
align at the boundary $r=\mathcal{R}$. Since we are dealing with a
static, non-rotating, spherically symmetric spacetime, the outer
region is represented by the de Sitter Schwarzschild vacuum
solution, given by the following line element
\begin{equation}\label{23}
ds^2_+ = \bigg(1 - \frac{2 M}{r} + \frac{\Lambda r^2}{3}\bigg) dt^2
- \bigg(1 - \frac{2 M}{r} + \frac{\Lambda r^2}{3}\bigg)^{-1} dr^2 -
r^2 \big(d\theta^2 + \sin^2 \theta \, d\phi^2\big).
\end{equation}
Here, plus sign indicates the exterior spacetime and $M$ is the
total mass contained within the boundary of the compact star for the
inner region. Substituting the value from Eq.\eqref{15} into
\eqref{7}, we have
\begin{equation}\label{24}
ds^2_- = e^{\big(\xi +\frac{1}{2} r \varphi  \sqrt{r^2 \chi
}\big)^2} dt^2 - e^{(r^2 \chi +1)} dr^2 - r^2 \big(d\theta^2 +
\sin^2 \theta \, d\phi^2\big).
\end{equation}
Since the cosmological constant is extremely small ($\Lambda=1.1056
10^{-46}km^{-2}$) \cite{As}, we can ignore it in our calculations.
to connect the star internal spacetime with the external spacetime
at the boundary, we must ensure that $g_{tt}$ and $g_{rr}$ and
$\frac{\partial}{\partial r} (g_{tt})$ are continuous across the
boundary $r=\mathcal{R}$. This requirement provides a set of
equations as shown below. This approach has been applied in previous
studies \cite{60}. We identify these constants specific to the
Schwarzschild solution by matching the temporal and radial
components of the metric from Eqs.\eqref{23} and \eqref{24}. We can
conclude that
\begin{eqnarray}\label{25}
1-\frac{2 M}{\mathcal{R}}&=&\bigg(\xi +\frac{1}{2} \mathcal{R}
\varphi \sqrt{\mathcal{R}^2 \chi }\bigg)^2,\\\label{38}
\bigg(1-\frac{2 M}{\mathcal{R}}\bigg)^{-1}&=&\mathcal{R}^2 \chi +1.
\end{eqnarray}
Differentiating Eq.\eqref{25} with respect to $\mathcal{R}$, it
follows that
\begin{eqnarray}\label{26}
\frac{2 M}{\mathcal{R}^2}&=&\frac{1}{2} \varphi \bigg(\mathcal{R}^3
\varphi \chi +2 \xi \sqrt{\mathcal{R}^2 \chi }\bigg).
\end{eqnarray}
We solve these equations together to find expressions for $\chi$,
$\xi$ and $\varphi$ as follows
\begin{eqnarray}\label{40}
\chi &=&\frac{2 M}{\mathcal{R}^2 (\mathcal{R}-2 M)}, \\\label{41}
\xi&=&\frac{\sqrt{\frac{M}{\mathcal{R}-2 M}} (2\mathcal{R}-5 M)}{2
\sqrt{M} \sqrt{\mathcal{R}}},\\\label{42}
\varphi&=&\frac{\sqrt{M}}{\sqrt{2} \mathcal{R}^{3/2}}.
\end{eqnarray}

Following O'Brien and Synge's method \cite{61} for an anisotropic
fluid sphere, the radial pressure must be continuous at the boundary
of the star, i.e., $p_\mathbf{r}(\mathcal{R}-0) =
p_\mathbf{r}(\mathcal{R}+0)$. Since there is no matter outside the
stellar surface, the radial pressure just beyond the boundary
vanishes, implying $p_\mathbf{r}(\mathcal{R}+ 0)$. Consequently, we
set $p_\mathbf{r}(\mathcal{R}- 0)$ at the surface. The full form of
the resulting boundary condition, obtained by substituting the
expressions for $B_1$ and $B_2$ into Eq.\eqref{16}, is quite lengthy
and has therefore been relocated to Appendix \textbf{C} for
completeness. This expression governs the behavior of the system
under the modified equation of state and plays a fundamental role in
the structure equations developed in the subsequent sections.

The constants in the metric coefficients for various compact stars
(EXO 1785-248 \cite{62}, Vela X-1 \cite{63}, SMC X-4 \cite{63}, LMC
X-4 \cite{63}, 4U 1538-52 \cite{63}) are presented in Table
\textbf{1}. In this table, OM denotes the observed mass, OR
represents the observed radius, EM indicates the estimated mass and
ER corresponds to the estimated radius.
\begin{table}\caption{Values of $\chi(km^{-2})$,
$\xi(km^{-2})$ and $\varphi$ for various compact stars.}
\begin{center}
\begin{tabular}{|c|c|c|c|c|c|c|c|}
\hline  Star & OM $\mathcal{M_{\odot}}$ & OR km & EM & ER& $\chi$&
$\xi $& $\varphi$
\\
\hline EXO 1785-248& $1.3\pm0.2$ & $8.85\pm0.4$ & 1.2 & 9.2 & 0.0063
& 0.0050 & 0.7790
\\
Vela X-1& $1.77\pm0.08$ & $9.56 \pm0.08$ & 1.77
&9.5&0.0106&0.0112&1.9802
\\
SMC X-4 &  $1.29\pm0.05$ & $8.831\pm0.09$ & 1.29
&8.8&0.0082&0.0057&1.2413
\\
LMC X-4 & $1.04\pm0.09$ & $8.301\pm0.2$ & 1.04 &8.3&0.0073&0.0046&
0.7625
\\
4U 1538-52&$0.87\pm0.07$ & $7.866\pm0.21$ & 0.87 &7.8&0.0072&
0.0042&0.6129
\\
\hline
\end{tabular}
\end{center}
\end{table}

\section{Physical Analysis}

In this section, we investigate the viability of hybrid stars by
examining the key physical properties of EXO 1785-248. This star has
been chosen because it serves as an excellent example of a hybrid
star candidate. Its intense X-ray emissions and unique binary
dynamics make it stand out, with a perfect balance of features like
matter transfer and gravitational interactions. Additionally, its
graphical representation highlights the suitability of the model for
studying such systems. These properties include metric components,
$\rho$, $p_\mathbf{r}$, $p_t$ and anisotropy, gradients of $\rho$,
$p_\mathbf{r}$, $p_t$, density and pressure of quark matter, EoS
parameter, energy conditions, mass-radius function. For all
graphical analyses, we use the parameter values $\zeta = 2, 2.2,
2.4, 2.6, 2.8$, $k=8\pi$ and $\mathcal{B}_g=67 MeV/fm^3$ to
illustrate the results.

\subsection{Metric Components}

Metric potentials are key elements in the geometry of spacetime in
GR and MTGs. They are essential for modeling dense astrophysical
objects. Figure \textbf{1} shows that both metric potentials,
$e^{\alpha}$ and $e^{\beta}$, are plotted against $r$. This
indicates that both potentials increase steadily with $r$.
\begin{figure}\center
\epsfig{file=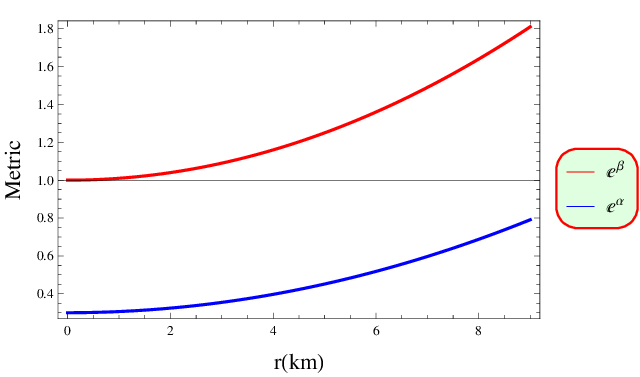,width=.55\linewidth}\caption{Graph of Metric
versus $r$.}
\end{figure}

\subsection{Behavior of $\rho$, $p_\mathbf{r}$, $p_t$ and Anisotropy}

For our model to be physically plausible, the density $\rho$, radial
pressure $p_\mathbf{r}$, and tangential pressure $p_t$ must remain
positive throughout the interior of the fluid sphere. These values
should also decrease steadily with increasing radius and be finite
at the star's boundary. To verify this, we analyze the graphical
behavior of density and pressure in Figure \textbf{2}. This shows
that both density and pressures are positive within the star, with
no physical or geometric singularities in the model. Figure
\textbf{3} illustrates how radial and transverse pressures relate to
density. It shows that both pressures increase linearly with
density, depending on different values of $\zeta$. The pressure
anisotropy, defined as $\Delta = p_t - p_\mathbf{r}$, may be
positive (when $p_t > p_\mathbf{r}$) or negative (when $p_t <
p_\mathbf{r}$). Anisotropy is significant in compact star models.
Figure \textbf{4} reveals that in this model, radial pressure
exceeds tangential pressure, resulting in negative anisotropy, which
creates an inward-directed attractive force.
\begin{figure}\center
\epsfig{file=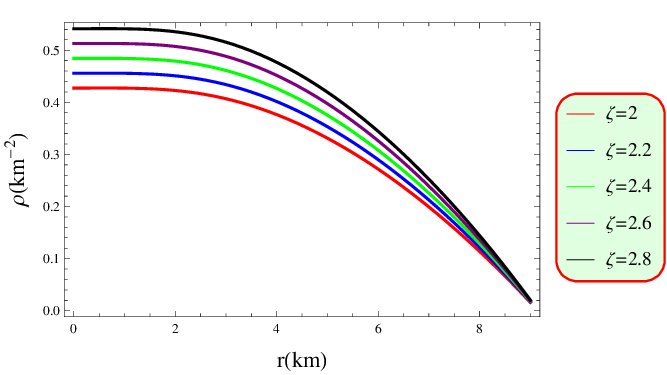,width=.55\linewidth}\epsfig{file=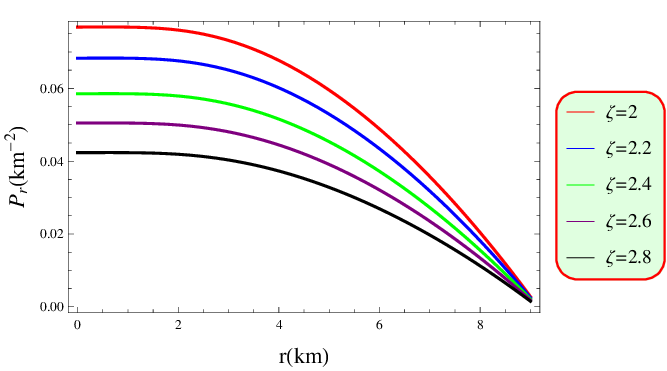,width=.55\linewidth}
\epsfig{file=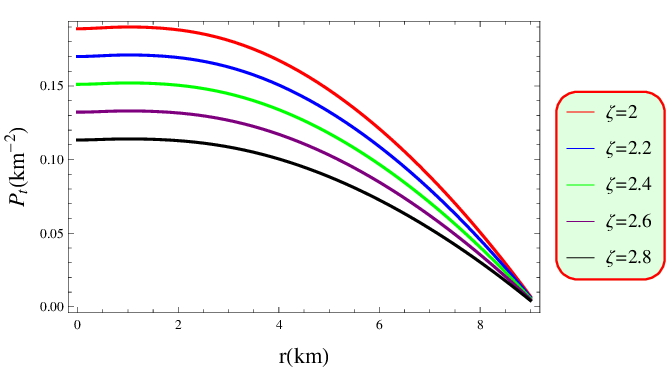,width=.55\linewidth}\caption{Plots of $\rho$,
$p_\mathbf{r}$ and $p_t$ against $r$.}
\end{figure}
\begin{figure}\center
\epsfig{file=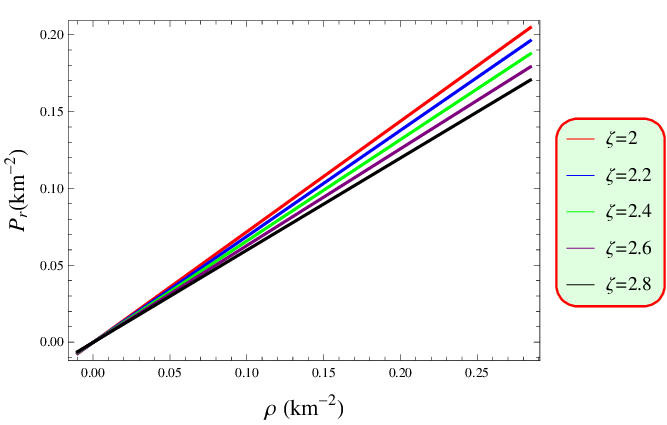,width=.55\linewidth}\epsfig{file=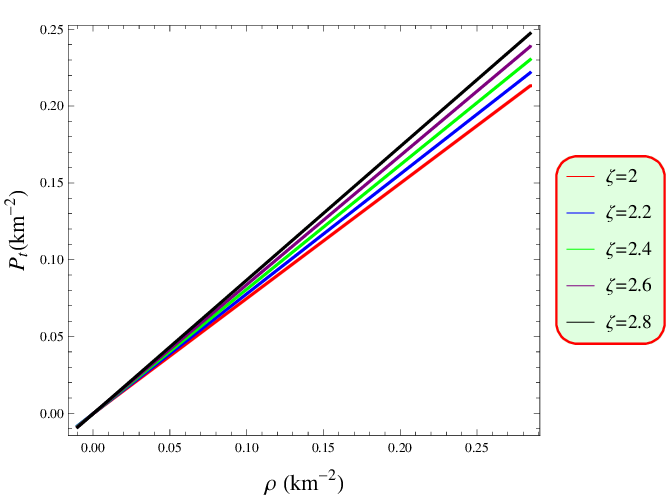,width=.47\linewidth}
\caption{Graphs of relationship between $p_\mathbf{r}$ and $p_t$
with $\rho$.}
\end{figure}
\begin{figure}\center
\epsfig{file=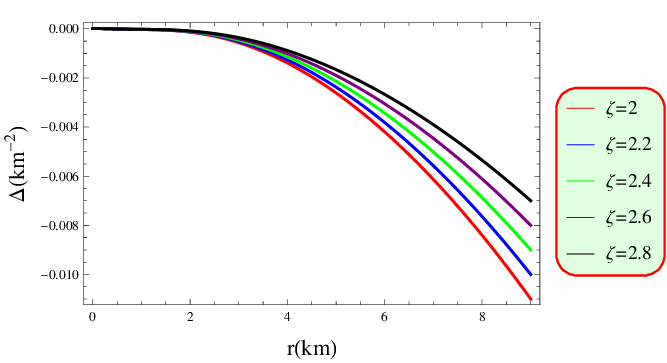,width=.55\linewidth}\caption{Plot of $\Delta$
against $r$.}
\end{figure}

\subsection{Gradients}

To determine the optimal values of $\rho$, $p_\mathbf{r}$ and $p_t$
within the star's interior, we examine the gradients of pressure and
density through plotted data. As shown in Figure \textbf{5}, these
gradients are uniformly negative across the star's interior. This
pattern suggests that both energy density and pressure are maximized
at the core and progressively diminish towards the outer layers.
\begin{figure}\center
\epsfig{file=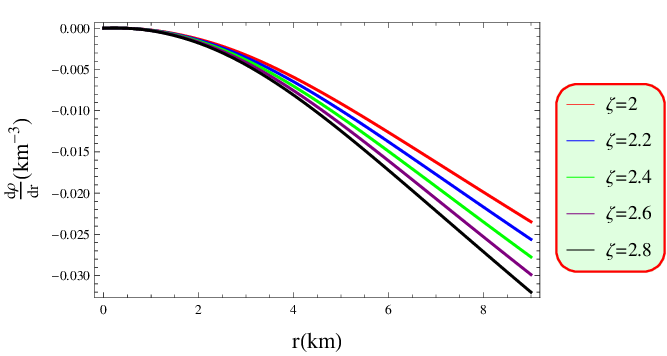,width=.55\linewidth}\epsfig{file=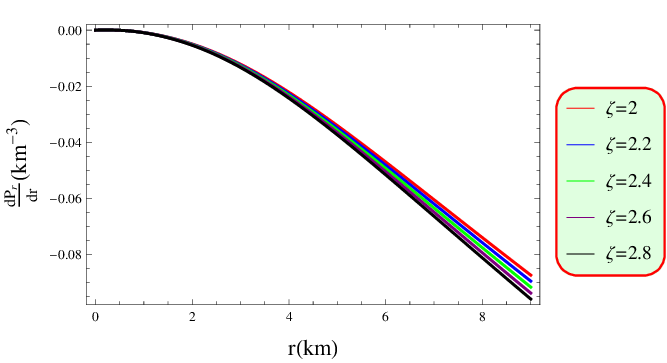,width=.55\linewidth}
\epsfig{file=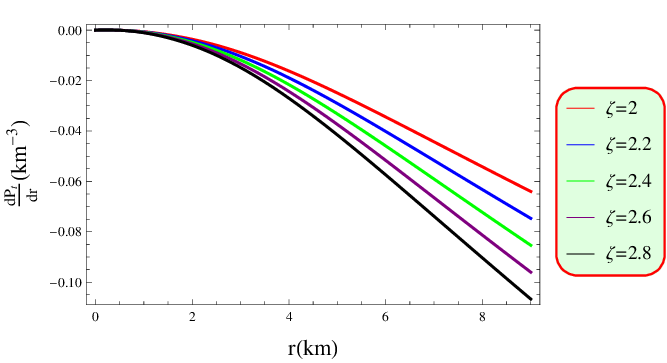,width=.55\linewidth}\caption{Plots of
$\frac{d\rho}{dr}$, $\frac{dp_\mathbf{r}}{dr}$ and $\frac{dp_t}{dr}$
against $r$.}
\end{figure}

\subsection{Density and Pressure of Quark Matter}

In quark matter, density is extremely high, often exceeding nuclear
densities found in atomic nuclei as quarks are packed closely within
a deconfined state. The pressure is likewise immense, driven by
interactions between quarks and it plays a crucial role in
supporting structures like neutron stars against gravitational
collapse. Figure \textbf{6} shows how the $\rho_\mathbf{q}$ and
$p_\mathbf{q}$, which are influenced by quark matter, change within
the star. Both values are positive for each coupling parameter
$\zeta$ given in the figure. They start high inside the star and
gradually decrease towards the boundary.
\begin{figure}\center
\epsfig{file=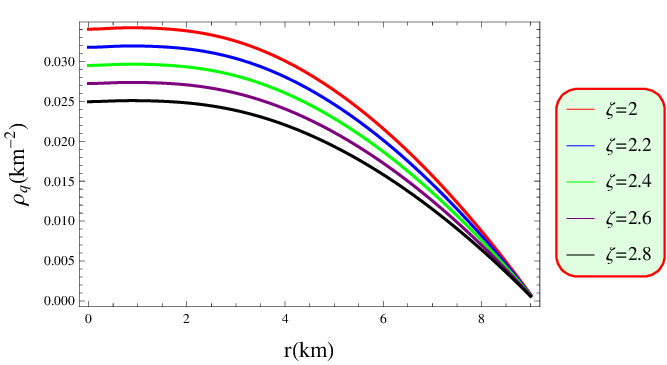,width=.55\linewidth}\epsfig{file=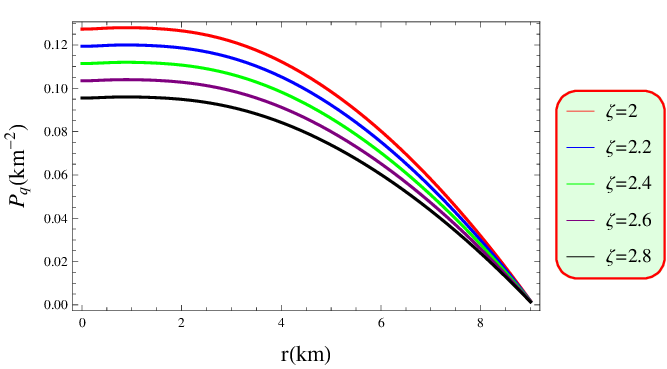,width=.55\linewidth}
\caption{Plots of $\rho_\mathbf{q}$ and $p_\mathbf{q}$ against $r$.}
\end{figure}

\subsection{Equation of State Parameter}

In our model, the EoS parameters $\omega_r$ and $\omega_t$ defined
by the relations $\omega_r = \frac{p_\mathbf{r}}{\rho}$ and
$\omega_t = \frac{p_t}{\rho}$, respectively, must lie within the
range $(0, 1)$. In astrophysics and cosmology, these parameters play
a vital role in describing the behavior of various components in the
universe. Figure \textbf{7} represents the behavior of $\omega_r$
and $\omega_t$, both of which lie within this interval. This range
suggests that the matter in our system is conventional, non-exotic
nature.
\begin{figure}\center
\epsfig{file=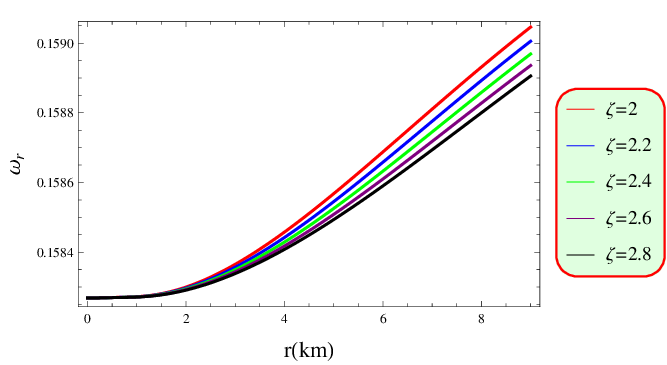,width=.55\linewidth}\epsfig{file=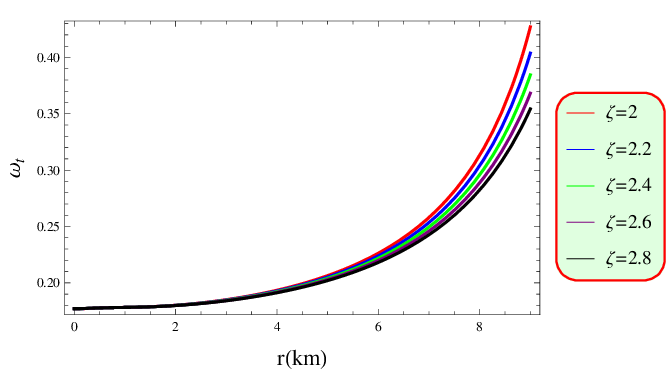,width=.55\linewidth}
\caption{Graphs of EoS against $r$.}
\end{figure}

\subsection{Energy Conditions}

To ensure realistic matter and energy distributions, certain
conditions must link energy density  and pressures, especially in
GR. These energy conditions help to classify matter as normal or
exotic and are defined by specific mathematical inequalities.
\begin{enumerate}
\item Null energy condition is represented as

$0 \leq \rho + p_\mathbf{r}, \quad 0 \leq \rho + p_t$.
\item Dominant energy condition is expressed as

$0 \leq \rho - p_\mathbf{r}, \quad 0 \leq \rho - p_t$.
\item Weak energy condition is presented as

$ 0 \leq \rho + p_\mathbf{r}, \quad 0 \leq \rho + p_t, \quad 0 \leq
\rho$.
\item Strong energy condition is stated as

$ 0 \leq \rho + p_\mathbf{r}, \quad 0 \leq \rho + p_t, \quad 0 \leq
\rho + p_\mathbf{r} + 2p_t$.
\end{enumerate}
These inequalities are plotted in Figure \textbf{8} to check if they
hold throughout the star's interior. The plots show that all these
conditions are met across the entire star, confirming that the
matter distribution is normal (not exotic).
\begin{figure}\center
\epsfig{file=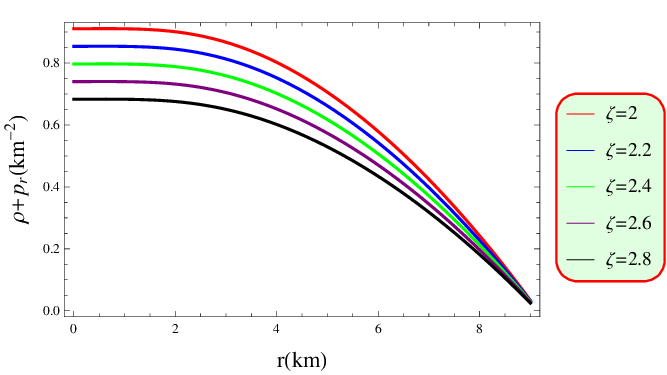,width=.55\linewidth}\epsfig{file=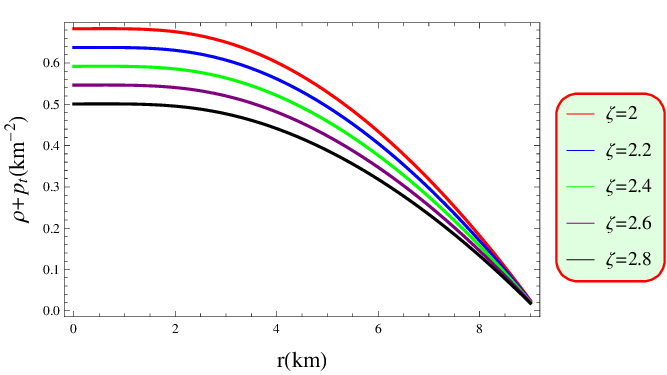,width=.55\linewidth}
\epsfig{file=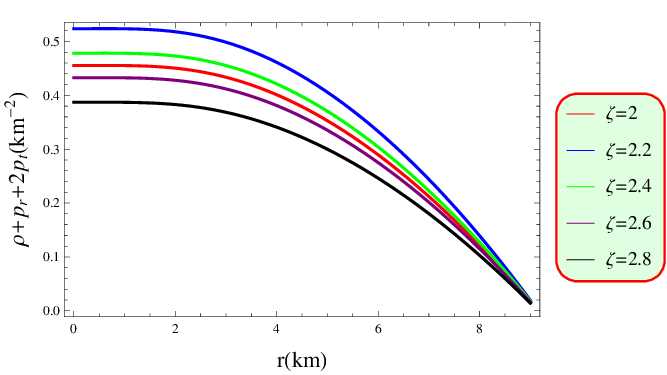,width=.55\linewidth}\epsfig{file=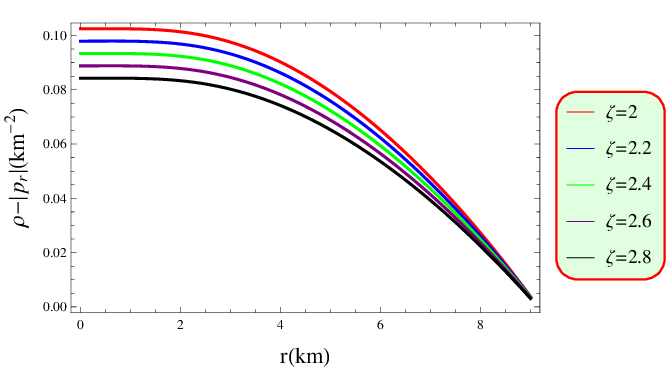,width=.55\linewidth}
\epsfig{file=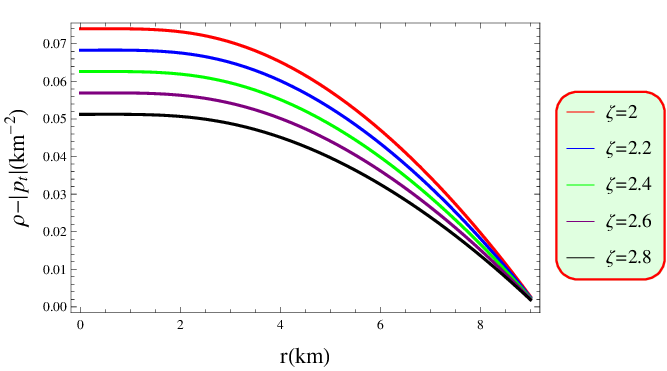,width=.55\linewidth} \caption{Graphs of energy
conditions against $r$.}
\end{figure}

\subsection{Mass-Radius Function}

The radius of the star is determined by the condition $p(R) = 0$.
The gravitational mass ($M(R)$) and baryon mass ($M_B$) \cite{es} of
a compact star differ fundamentally in both their definitions and
physical interpretations. Gravitational mass is defined as $M =
\int_0^R 4\pi r^2 \rho , dr$, where $\rho$ represents the total
energy density, including contributions from rest mass, internal
energy and pressure. This mass generates the star's gravitational
field and accounts for energy losses such as radiation or neutrino
emissions, thus reflecting the effective mass of the star. In
contrast, baryon mass is defined as $M_B = \int_0^R 4\pi r^2
e^{\beta/2} \rho dr$, where $e^{\beta/2}$ is a factor that
incorporates the effects of spacetime. It quantifies the total rest
mass of baryonic matter, such as protons and neutrons, without
considering energy losses. As a result, the baryon mass is always
greater than the gravitational mass.

Figure \textbf{9} shows that both masses increase with the surface
radius $R$. The gravitational mass typically falls within values
such as $0.1, 0.2$ etc., while the baryon mass is significantly
higher, with values like $1, 2, 3$ and so on. This difference arises
because the baryon mass includes the factor $e^{\beta/2}$,
amplifying the effects of spacetime. The distinction between these
two masses is crucial for understanding the physical and
gravitational properties of hybrid stars. The plotted results align
with theoretical expectations, emphasizing the significant role of
spacetime effects and relativistic contributions in differentiating
baryonic and gravitational masses.
\begin{figure}\center
\epsfig{file=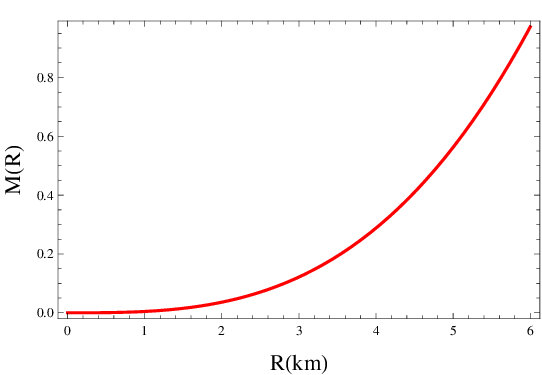,width=.5\linewidth}\epsfig{file=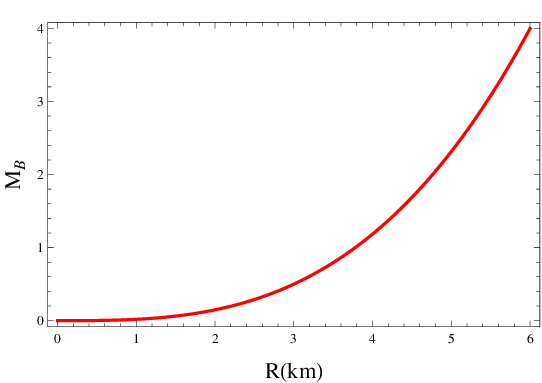,width=.5\linewidth}
\caption{Plots of $M(R)$ and $M_{B}$ against $R$.}
\end{figure}

\section{Stability}

In this section, we examine the stability by analyzing the
compactness, redshift, causality condition and Herrera cracking as
well as adiabatic index.

The compactness function, denoted as $U(R) =\frac{M(R)}{R}$, is a
key parameter for assessing a star's stability. To avoid
gravitational collapse, a star's compactness should stay below the
threshold value of $\frac{4}{9}$ \cite{1a3}. This ensures that the
gravitational pull is not strong enough to trigger collapse. Thus,
maintaining $U(R)$ below this limit is essential for a stable
configuration. Figure \textbf{10} (left plot) shows that the
compactness steadily increases but remains safely within the
required stability range.
\begin{figure}\center
\epsfig{file=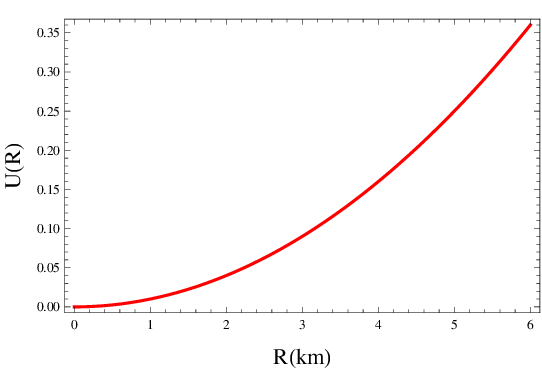,width=.5\linewidth}\epsfig{file=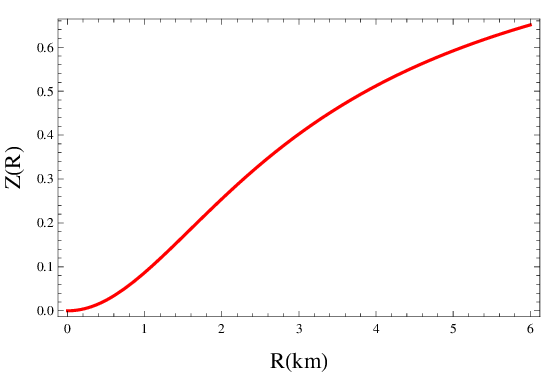,width=.5\linewidth}
\caption{Plots of $U(R)$ and $Z(R)$ against $R$.}
\end{figure}

The redshift parameter is essential for analyzing the stability of
gravastars and identifying them through their structural
characteristics. For an isotropic matter distribution, the redshift
typically remains below a value of 2 \cite{hy}. It can be calculated
using the formula $ Z(R) = \big(1 - 2U(R)\big)^{-\frac{1}{2}} - 1$.
As shown in Figure \textbf{10} (right plot), the redshift parameter
consistently remains within the required limit, $Z(R) < 2$. Within
the star, the redshift stays positive and finite, gradually
increasing as it approaches the surface, thereby providing insight
into the stability of our model.

The speed of sound is an important factor that must be kept within a
specific range $[0, 1]$ to ensure that the model remains stable
\cite{64}. The square of the radial and tangential sound speeds can
be determined using the following expressions
\begin{equation}\label{1a}
v_r^2= \frac{d{p}_{r}}{d \rho},\quad v_t^2= \frac{d {p}_{t}}{d
\rho}.
\end{equation}
According to the causality conditions, for the model to be stable
within the star's interior, the following conditions must be hold:
$0 \leq v_r^2 \leq 1$ and $0 \leq v_t^2 \leq 1$. Additionally, the
radial and tangential components of sound speed are important in
testing cracking instability, as described in criteria by Herrera
\cite{65}. The cracking condition can be expressed mathematically as
$0 \leq \mid v_t^2 - v_r^2 \mid \leq 1$. These values remain below 1
across all variations of parameter $\zeta$ in our model. Figure
\textbf{11} confirms that the model aligns with the stability
conditions specified above.
\begin{figure}\center
\epsfig{file=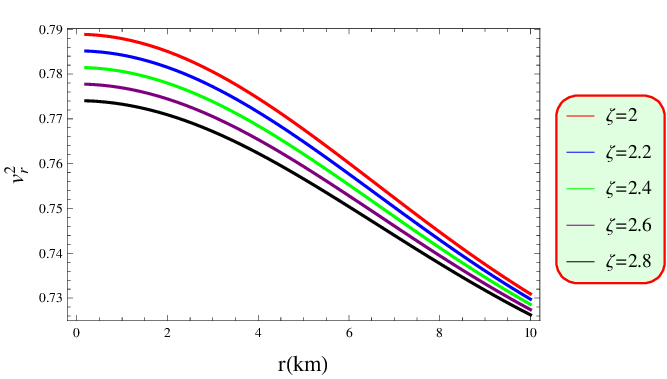,width=.55\linewidth}\epsfig{file=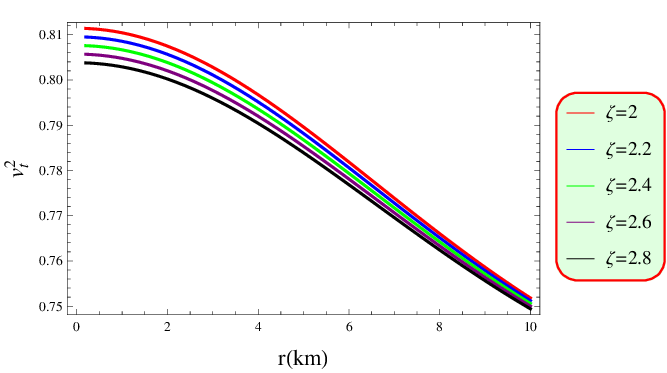,width=.55\linewidth}
\epsfig{file=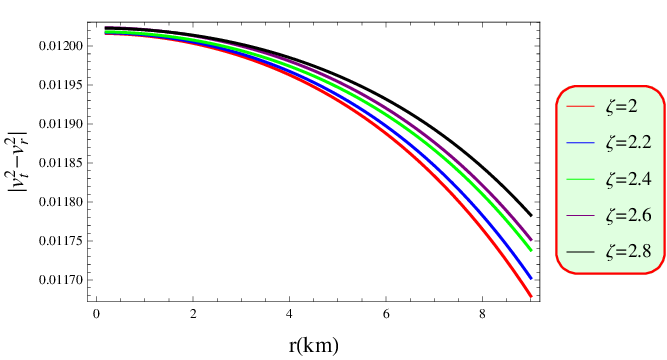,width=.55\linewidth} \caption{Plots of
causality conditions and Herrera cracking against $r$.}
\end{figure}

The adiabatic index $\Gamma$  is an important astrophysical
parameter that indicates the stability of a fluid. The radial
($\Gamma_r$) and tangential ($\Gamma_t$) components of the adiabatic
index are expressed as
\begin{equation}\nonumber
\Gamma_{r}=\frac{v_r^2 (p_\mathbf{r}+\rho )}{p_\mathbf{r}},\quad
\Gamma_{t}=\frac{v_t^2(p_{t}+\rho)}{p_{t}}.
\end{equation}
In the context of compact objects, $\Gamma_r$ and $\Gamma_t$ must be
greater than $\frac{4}{3}$ to ensure a stable configuration, as this
condition prevents collapse due to gravitational forces \cite{66}.
The adiabatic index thus plays a critical role in analyzing the
stability of stellar structures and understanding the behavior of
matter under extreme conditions. Figure \textbf{12} demonstrates
that the system achieves stability by satisfying the required limit.
This stability supports the formation of a viable hybrid star,
capable of with standing gravitational collapse and preventing
energy dissipation.
\begin{figure}\center
\epsfig{file=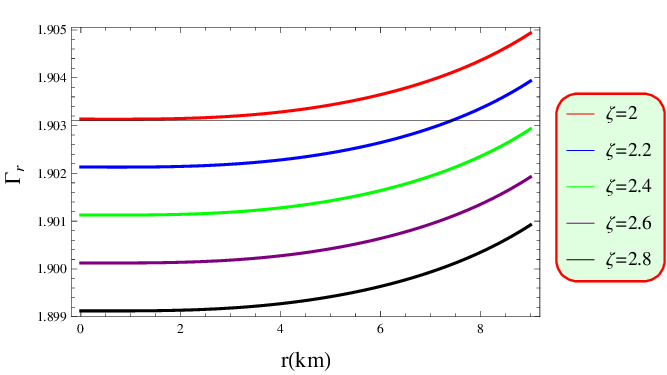,width=.55\linewidth}\epsfig{file=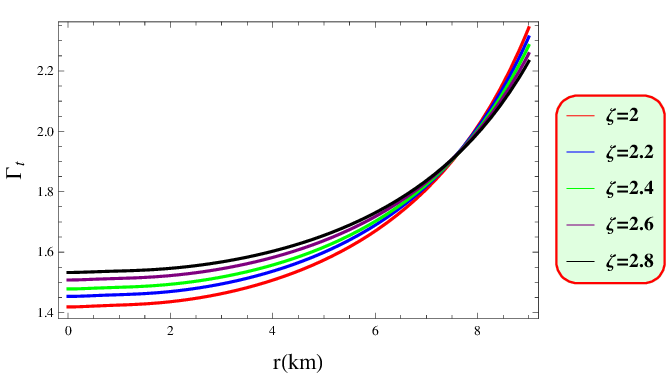,width=.55\linewidth}
\caption{Plots of $\Gamma_r$ and $\Gamma_t$ against $r$.}
\end{figure}
\begin{figure}\center
\epsfig{file=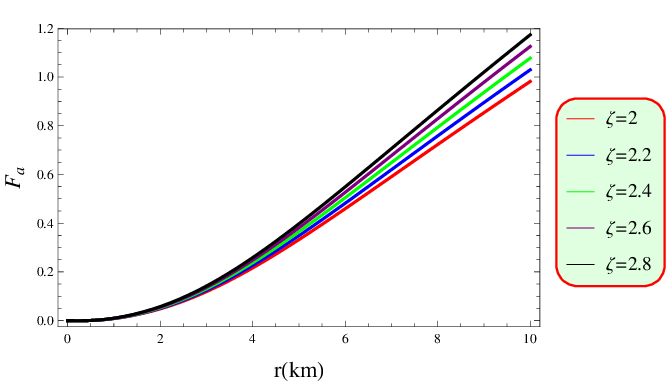,width=.55\linewidth}\epsfig{file=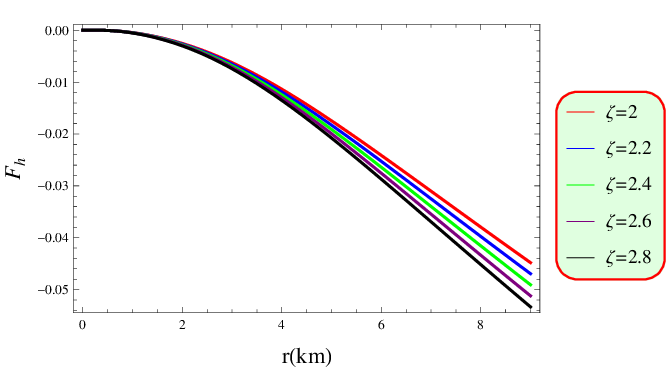,width=.55\linewidth}
\epsfig{file=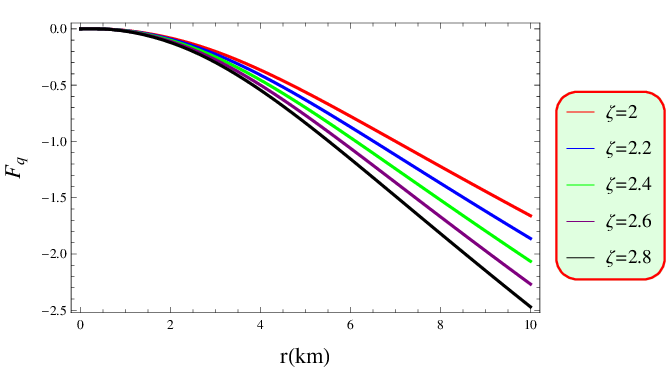,width=.55\linewidth}\epsfig{file=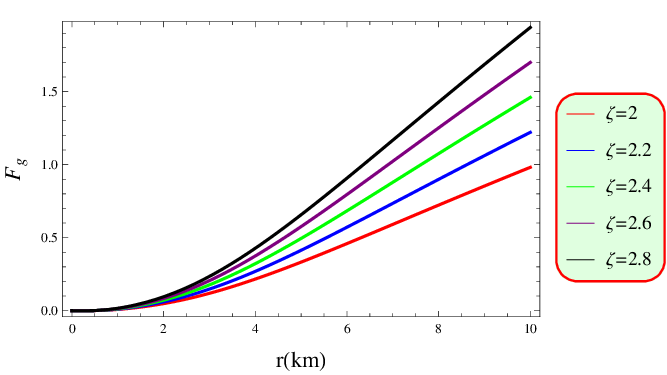,width=.55\linewidth}
\caption{Graphs of TOV equation versus $r$.}
\end{figure}

\subsection{Equilibrium Equations}

In this subsection, the Tolman-Oppenheimer-Volkoff (TOV) equation
has been modified to incorporate the effects of $f(Q)$ gravity and
quark matter \cite{50a}. The equation represents the equilibrium of
the total forces acting on the system, including hydrostatic,
gravitational, anisotropic and quark matter contributions. This
balance is expressed as
\begin{equation}\label{h21}
F_g + F_h + F_a + F_\mathbf{q} = 0.
\end{equation}
Here, the mathematical expressions for each force are as follows
\begin{itemize}
\item Gravitational Force: $ F_g = -\frac{\alpha^{\prime}}{2} (\rho + p_\mathbf{r}).$
\item Hydrostatic Force: $F_h = -\frac{dp_\mathbf{r}}{dr}.$
\item Anisotropic Force: $F_a = \frac{2}{r} (p_t - p_\mathbf{r}).$
\item Quark Matter Force: $F_\mathbf{q} = -\frac{\alpha^{\prime}}{2}
(\rho_\mathbf{q} + p_\mathbf{q}) - \frac{dp_\mathbf{q}}{dr}.$
\end{itemize}
The modified TOV equation is then written as
\begin{equation}\label{h22}
-\frac{\alpha^{\prime}}{2} (\rho + p_\mathbf{r}) -
\frac{dp_\mathbf{r}}{dr} + \frac{2}{r} (p_t - p_\mathbf{r}) -
\frac{\alpha^{\prime}}{2} (\rho_\mathbf{q} + p_\mathbf{q}) -
\frac{dp_\mathbf{q}}{dr} = 0.
\end{equation}
To ensure the model's stable equilibrium, these forces must be
properly balanced. Figure \textbf{13} illustrates that stability is
maintained when the sum of $F_{a}$, $F_{g}$, $F_{h}$ and
$F_{\mathbf{q}}$ is zero. This indicates that the hybrid star
remains in equilibrium for various values of $\zeta$.

\subsection{Bounds of the Coupling Parameter}

We now aim to determine the bounds of the coupling parameter
$\zeta$. Using the trace energy condition, $\rho - p_\mathbf{r} -
2p_t > 0$, at the star surface, we determine the following limit for
$\zeta$. Additionally, since the transverse sound velocity inside
the star must be less than 1 (i.e., $ \frac{dp_t}{d\rho} < 1$ at the
boundary), this condition also gives us another limit for $\zeta$.
The mathematical expression can be obtained in Appendix \textbf{D}.
By combining the two inequalities for $\zeta$ above, we obtain a
reasonable bound for $\zeta$ as $\Psi< \zeta< 1 +\Phi$. The final
bound ensures that the model remains physically consistent with both
energy and stability conditions.

\section{Summary and Discussion}

In this paper, we have developed a model of anisotropic compact
stars, specifically a hybrid star, in which strange quark matter and
normal hadronic matter are combined using a two-fluid distribution.
This model is free from any mathematical or geometrical issues. We
have investigated the solution of the hybrid star within the
framework of $f(Q)$ gravity, utilizing the Finch-Skea metric. For
our analysis, we have selected the compact star EXO 1785-248, which
serves as a strong example of a hybrid star. The key physical
properties and stability of this hybrid star have been examined and
are summarized as follows.
\begin{itemize}
\item The metric potentials are continuous and show a monotonic increase
with $r$. This behavior reflects a smooth evolution of the spacetime
structure. The absence of abrupt or sudden changes from the star's
center to its boundary confirms a singularity-free and physically
consistent model. The finite nature of these potentials at the core
validates the theoretical framework, demonstrating a realistic
stellar configuration that adheres to fundamental physical and
geometric constraints. This makes the model highly suitable for
describing compact stars (Figure \textbf{1}).
\item The physical behavior of the hybrid star model shows that
$\rho$, $p_\mathbf{r}$, $p_t$, $\rho_\mathbf{q}$, and $p_\mathbf{q}$
reach their maximum values at the center and gradually decrease
towards the surface. This indicates a well-behaved and physically
consistent solution. As the parameter $\zeta$ increases, the density
and both radial as well as transverse pressures decrease. This
suggests that $\zeta$ has a significant impact on the star's
internal structure, potentially influencing its overall equilibrium
and matter distribution (Figures \textbf{2} and \textbf{6}).
\item We have noticed that radial and transverse pressures change
in a way that closely follows a linear pattern with density, and
this happens for all values of $\zeta$. This shows that the
pressures adjust smoothly, helping the star to maintain a proper
balance in its structure. If the pressures were uneven, the star
might either collapse or become too spread out (Figure \textbf{3}).
\item We have analyzed that the anisotropic factor is negative which
satisfies the condition $p_t< p_\mathbf{r}$. This means that the
force from anisotropy combines with the gravitational force to help
to compress the star (Figure \textbf{4}).
\item It is found that the gradients of $\rho$, $p_\mathbf{r}$, and $p_t$ are
negative throughout the star interior. This confirms that these
quantities are the highest at the core of the star. As a result, the
negative gradients support a stronger gravitational pull, helping
the star to compress further (Figure \textbf{5}).
\item The EoS parameter between 0 and 1 indicates a consistent
relationship between pressure and energy density. This confirms that
the hybrid star meets the required physical conditions (Figure
\textbf{7}).
\item  All energy conditions have positive results, showing that
the energy density and pressures follow acceptable physical limits.
This confirms the presence of hadronic matter in the hybrid star
(Figure \textbf{8}).
\item The behavior of the mass function shows an increasing trend as the radial
coordinate increases and the mass distribution shows a physically
possible and realistic behavior for the hybrid star (Figure
\textbf{9}).
\item We have also checked the stability of the gravastar by examining
compactness, redshift, causality conditions and the adiabatic index.
Our findings confirm that the stability limits support the existence
of a stable hybrid star (Figure \textbf{10-12}).
\item We have observed that $F_a$ and $F_g$ have positive values,
whereas $F_{\mathbf{q}}$ and $F_h$ are negative. Since the total
force balance is zero, the system stays in equilibrium within this
framework (Figure \textbf{13}).
\end{itemize}

It is worth mentioning that compact stars such as neutron stars,
usually rotate and possess angular momentum. Therefore, a fully
spherically symmetric solution may not always be applicable.
However, for simplicity, this paper focuses on non-rotating compact
stars, excluding angular momentum. Previous studies in $f(Q)$
gravity have similarly analyzed non-rotating compact stars
\cite{66a}. Within the framework of $f(Q)$ gravity, we have explored
the Finch-Skea metric using two fluid distributions: strange quark
matter and hadronic matter. Our analysis of a compact star candidate
in $f(Q)$ gravity has demonstrated the viability of addressing both
central singularity and stability conditions for the internal fluid
distribution. This led us to conclude that such a dense, compact
object made of quark matter could serve as the excellent model for
hybrid stars, allowing the investigation of their physical
properties on both theoretical and astrophysical scales.
Furthermore, we note that, beyond the conventional GR approach,
other higher-order gravity theories can also be employed to explore
these physical characteristics under specific constraints.

In contrast, $f(Q)$ gravity \cite{66a} incorporates non-metricity
$Q$ as a fundamental component of gravity, making it uniquely suited
to model environments where anisotropic pressures and non-singular
configurations exist. The theory provides a more flexible approach
for handling the internal structure of hybrid stars, which often
involve extreme densities and complex phase transitions between
quark and hadronic matter. Studies have shown that $f(Q)$ gravity
can support supermassive hybrid stars and explain the observed
mass-radius relationships for compact stars more effectively than
$f(R)$ and $f(\mathcal{T})$. Moreover, $f(Q)$ gravity offers
non-singular solutions that can accommodate exotic matter states
without the singularities typically associated with $f(R)$ gravity.
This flexibility is crucial for modeling hybrid stars, where the
internal pressure distribution can vary significantly. This capacity
to model smooth, continuous solutions is essential for understanding
the complex internal structures of these stars.

While $f(Q)$ gravity is a relatively new theory with limited
empirical validation compared to well-established frameworks like
$f(R)$, initial results have shown promising potential, particularly
in the modeling of non-singular hybrid stars and the handling of
anisotropic pressures. Comparative studies with $f(R)$ and
$f(\mathcal{T})$ are essential to fully explore the capabilities of
$f(Q)$ and identify areas where it could be further refined. As
research progresses, $f(Q)$ gravity has the potential to become a
viable alternative to $f(R)$, offering new insights into the
structure of compact stars and other astrophysical phenomena.

We have found that the $f(Q)$ theory is highly flexible in modeling
gravitational interactions, particularly in environments
characterized by anisotropic pressures and non-singular
configurations. Bhar et al. \cite{67} employed the Krori-Barua
metric to model hybrid stars within the framework of $f(Q)$ gravity,
while we adopted the Finch-Skea metric, which offers a more flexible
and realistic representation of compact stars, especially in
evaluating stability. Quantitatively, our Herrera cracking graph
exhibited a monotonic decrease across the stellar interior,
indicating smooth and stable behavior throughout. In contrast, the
analysis \cite{67} showed a cracking profile that initially
increased and then decreased, suggesting fluctuating stability and
possible local instabilities. Our model also satisfies the condition
$\Gamma > \frac{4}{3}$ for the adiabatic index at all radii,
ensuring dynamical stability, whereas this index in \cite{67} varied
and dropped below this threshold in certain regions. The maximum
compactness factor in our configuration reaches $U(R) = 0.192$,
remaining well within the Buchdahl limit, while the corresponding
values in \cite{67} appeared higher and more variable. Additionally,
the surface redshift in our case attains a maximum of $Z(R)=0.275$,
comfortably within the theoretical bound, in contrast to the results
\cite{67}, which approached or in some cases exceeded this limit.
These distinctions illustrated that the Finch-Skea metric yield a
more consistent stable and physically viable configuration compared
to the Krori-Barua metric, particularly with respect to the Herrera
cracking, sound speed criteria and adiabatic behavior. Our results
align well with the previous studies in $f(R)$ \cite{68} and
$f(\mathcal{T})$ \cite{66d} theories of gravity. We can conclude
that our study provides a comprehensive and stable hybrid star
solution in $f(Q)$ gravity, offering enhanced insights into the
stability analysis.

\section*{Appendix A: Non-Metricity Scalar and Superpotential Representation}
\renewcommand{\theequation}{A\arabic{equation}}
\setcounter{equation}{0}

The non-metricity scalar is defined by
\begin{equation}\label{2}
Q=-g^{\gamma\psi}(\mathbb{L}^{\mu}_{~\nu\gamma}\mathbb{L}^{\nu}_{~\psi\mu}
-\mathbb{L}^{\mu}_{~\nu\mu}\mathbb{L}^{\nu}_{~\gamma\psi}).
\end{equation}
Here, the Levi-Civita connection is represented by the deformation
tensor as $\Gamma^\mu_{\nu \varsigma} = -\mathbb{L}^\mu_{\; \nu
\varsigma}$, where
\begin{equation}\label{3}
\mathbb{L}^{\mu}_{\;\nu\varsigma}=-\frac{1}{2}g^{\mu\lambda}
(\nabla_{\varsigma}g_{\nu\lambda}+\nabla_{\nu}g_{\lambda\varsigma}
-\nabla_{\lambda}g_{\nu\varsigma}).
\end{equation}
The superpotential is expressed as
\begin{equation}\label{4}
\mathbb{P}^{\mu\psi\gamma}=\frac{1}{4}\big[Q^{\psi\mu\gamma}-Q^{\mu\psi\gamma}
+Q^{\psi\mu\gamma}+Q^{\gamma\mu\psi}-\tilde{Q}_{\mu}g^{\psi\gamma}
+Q^{\mu}g^{\psi\gamma}\big],
\end{equation}
where
\begin{equation}\nonumber
Q_{\mu}=Q^{~\psi}_{\mu~\psi},\quad
\tilde{Q}_{\mu}=Q^{\psi}_{~\mu\psi}.
\end{equation}
Thus, $Q$ can be further rewritten as \cite{54}
\begin{equation}\label{5}
Q=-Q_{\mu\gamma\psi}\mathbb{P}^{\mu\gamma\psi}=-\frac{1}{4}(-Q^{\mu\psi\rho}Q_{\mu\psi\rho}
+2Q^{\mu\psi\rho}Q_{\rho\mu\psi}-2Q^{\rho}\tilde{Q}_{\rho}+Q^{\rho}Q_{\rho}).
\end{equation}

\section*{Appendix B: Matter Variables and Quark Matter}
\renewcommand{\theequation}{B\arabic{equation}}
\setcounter{equation}{0}

The expressions for the matter density and pressures associated with
regular hadronic matter are as follows
\begin{eqnarray}\nonumber
\rho&=&\bigg[4 \xi  r^{11} \varphi  \chi ^5 (4 \mathcal{B}_{g}
\kappa+36 \zeta  \chi +3 h \kappa-42)+8 \xi  r^9 \varphi  \chi ^4 (8
\mathcal{B}_{g} \kappa+30 \zeta \chi +6 h \kappa\\\nonumber &+&4
\chi -27) +4\xi r^7 \varphi \chi ^3 (24 \mathcal{B}_{g} \kappa+\chi
(6 \zeta (4-5 \chi )+13)+18 h \kappa-30)-8 \xi \\\nonumber
&\times&r^5\varphi \chi ^2 (-8 \mathcal{B}_{g} \kappa+2 \chi  (9
\zeta \chi +1)-6 h \kappa+3)+4 \varphi \chi  (4 \mathcal{B}_{g}
\kappa+\chi (8 \zeta  \chi -15)\\\nonumber &+&3 h \kappa)\xi r^3+4
\big(r^2 \chi \big)^{3/2} \big(16 \mathcal{B}_{g} \kappa \xi ^2+\chi
\big(\xi ^2 (5-16 \zeta \chi )+24 \zeta \varphi ^2\big)+12 h \kappa
\xi ^2\\\nonumber &-&3 \varphi ^2\big)+r^2 \big(r^2 \chi \big)^{3/2}
\big(4 \mathcal{B}_{g} \kappa \big(24 \xi ^2 \chi +\varphi
^2\big)-\big(4 \xi ^2 (\chi (\zeta \chi -5)+3)+5 \varphi ^2
\\\nonumber &\times&(3-20 \zeta  \chi )\big)2 \chi+3 h \kappa \big(24 \xi
^2 \chi +\varphi ^2\big)\big)+r^2 \big(r^2 \chi \big)^{5/2} \big(4
\xi ^2 \chi (16 \mathcal{B}_{g} \kappa+12\zeta \\\nonumber &\times&
\chi +12 h \kappa+5 \chi -18)+\varphi ^2 (16 \mathcal{B}_{g}
\kappa+\chi (\zeta (\chi +2)-13)+12 h \kappa-12)\big)\\\nonumber
&-&4 \kappa \xi ^2 (4 \mathcal{B}_{g}+3 h) \sqrt{r^2 \chi }+r^{12}
\varphi ^2 \chi ^5 \sqrt{r^2 \chi } (4 \mathcal{B}_{g} \kappa+108
\zeta \chi +3 h \kappa-66)+r
\\\nonumber &\times&\chi ^3 \big(r^2 \chi \big)^{3/2} \big(\varphi ^2 (16
\mathcal{B}_{g} \kappa+252 \zeta \chi +12 h \kappa+11 \chi -90)-24
\xi ^2 \chi \big)+2 r^8 \chi ^3 \\\nonumber &\times&\sqrt{r^2 \chi }
\big(2 \xi ^2 \chi (4 \mathcal{B}_{g} \kappa+3 (4 \zeta \chi +h
\kappa-6))+\varphi ^2 (12 \mathcal{B}_{g} \kappa+\chi (-33 \zeta
\chi +96 \zeta \\\nonumber &+&8)+9 h \kappa-27)\big)-48 \xi r^{13}
\varphi \chi ^6-18 r^{14} \varphi ^2 \chi ^6 \sqrt{r^2 \chi }-24 \xi
r \varphi \chi \bigg]\bigg[ (3 s-1)\\\label{18} &\times&
\kappa\sqrt{r^2 \chi } \big(r^2 \chi +1\big)^4 \big(2 \xi +r \varphi
\sqrt{r^2 \chi }\big)^2\bigg]^{-1},\\
\nonumber p_{r}&=& \bigg[\sqrt{r^2 \chi } \big(h \kappa \big(240 r^2
\xi ^4 \varphi ^2 \big(r^2 \chi \big)^{3/2}+60 r^4 \xi ^2 \varphi ^4
\big(r^2 \chi \big)^{5/2}+64 \xi ^6 \sqrt{r^2 \chi
}+r^{12}\\\nonumber &\times& \varphi ^6 \chi ^3 \sqrt{r^2 \chi }+12
r^{11} \xi  \varphi ^5 \chi ^3+160 r^7 \xi ^3 \varphi ^3 \chi ^2+192
r^3 \xi ^5 \varphi  \chi \big) \big(\chi  r^2+1\big)^4+s \\\nonumber
&\times&\big(-64 \xi ^4 \big(-16 \mathcal{B}_{g}\kappa \xi ^2+\chi
(16 \zeta  \chi -5) \xi ^2+3 \varphi ^2 (5-8 \zeta \chi )\big)
\big(r^2 \chi \big)^{3/2}-32 \\\nonumber &\times&r^2 \xi ^4
\big(\chi \big(4 (\chi  (\zeta  \chi -5)+3) \xi ^2+3 \varphi ^2
(25-44 \zeta \chi )\big)-6 \mathcal{B}_{g}\kappa \big(8 \chi  \xi
^2+5 \varphi ^2\big)\big)\\\nonumber &\times& \big(r^2 \chi
\big)^{3/2}+16 r^2 \xi ^2 \big(4 \chi (16 \mathcal{B}_{g}\kappa+12
\zeta \chi +5 \chi -18) \xi ^4+3 \varphi ^2 \xi^2(80
\mathcal{B}_{g}\kappa+\chi
\\\nonumber &\times&(8 \zeta (2-15 \chi )-5)-20) +6 \varphi ^4 (24
\zeta \chi -5)\big) \big(r^2 \chi \big)^{5/2}+16 r^4 \xi ^2
\big(\mathcal{B}_{g}\kappa \big(16 \\\nonumber &\times&\chi ^2 \xi
^4+360 \varphi ^2 \chi \xi ^2+15 \varphi ^4\big)+\chi \big(24 \chi
(2 \zeta \chi -3) \xi ^4-6 \varphi ^2 \xi ^2(\chi  (53 \chi
\zeta-64 \zeta
\\\nonumber &-&30)+55)+\varphi ^4 (316 \zeta  \chi
-75)\big)\big) \big(r^2 \chi \big)^{5/2}-4 r^4 \big(96 \chi ^2 \xi
^6-12 \varphi ^2 \chi (80 \mathcal{B}_{g}\kappa\\\nonumber &+&268
\zeta \chi +35 \chi -210) \xi ^4+\varphi ^4 (-240
\mathcal{B}_{g}\kappa+\chi (32 \zeta (5 \chi -9)+105)+120)
\\\nonumber &\times&\xi ^2+3 \varphi ^6 (1-8 \zeta \chi )\big)
\big(r^2 \chi \big)^{7/2}-18 r^{22} \varphi ^6 \chi ^8 \sqrt{r^2
\chi }+256 \mathcal{B}_{g}\kappa \xi ^6 \sqrt{r^2 \chi }+2
\\\nonumber &\times&r^{20} \varphi ^6 \chi ^7 (2 \mathcal{B}_{g}\kappa+54
\zeta \chi -33) \sqrt{r^2 \chi }+r^{18} \varphi ^4 \chi ^6
\big(\varphi ^2 (16 \mathcal{B}_{g}\kappa+252 \zeta \chi +11
\\\nonumber &\times&\chi -90)-\xi^2 \chi \big) \sqrt{r^2 \chi }
+2 r^{16} \varphi ^4 \chi ^5 \big(12 \chi  (10\mathcal{B}_{g}
\kappa+158 \zeta  \chi -125) \xi ^2+(12 \\\nonumber &\times&
\mathcal{B}_{g}\kappa+\chi (-33 \chi \zeta +96 \zeta +8)-27)\varphi
^2\big) \sqrt{r^2 \chi }+r^{14} \varphi ^2 \chi ^4 \big(2400 \chi ^2
\xi ^4+12\\\nonumber &\times& \varphi ^2 \chi (80
\mathcal{B}_{g}\kappa+68 \zeta \chi +45 \chi -330) \xi ^2+\varphi ^4
(16 \mathcal{B}_{g}\kappa+\chi (24 \zeta (\chi +2)-13)\\\nonumber
&-&12)\big) \sqrt{r^2 \chi }+2 r^{12} \varphi ^2 \chi ^3 \big(2
\mathcal{B}_{g}\kappa \big(240 \chi ^2 \xi ^4+360 \varphi ^2 \chi
\xi ^2+\varphi ^4\big)+\chi \big(48 \chi  (78 \zeta \\\nonumber
&-&85) \xi ^4+4 \varphi ^2 (\chi  (105-319 \chi  \zeta +672 \zeta
)-285) \xi ^2+ \varphi ^4 (20 \zeta \chi -3)\big)\big) \sqrt{r^2
\chi }\\\nonumber &-&192 r^{21} \xi \varphi ^5 \chi ^8-384 r \xi ^5
\varphi  \chi +24 r^{19} \xi \varphi ^5 \chi ^7 (2
\mathcal{B}_{g}\kappa+42 \zeta \chi -29)+r^3 \xi ^5 \varphi \chi
\\\nonumber &\times&(12 \mathcal{B}_{g}\kappa+\chi  (8 \zeta  \chi -15))+24
r^{17} \xi \varphi ^3 \chi ^6 \big(\varphi ^2 (
\mathcal{B}_{g}\kappa+94 \zeta \chi +5 \chi -39)-\xi ^2 \\\nonumber
&\times&\chi 80 \big)+64 r^5 \xi ^3 \varphi \chi ^2 \big(\big(-68
\zeta  \chi ^2+6 \chi +48 \mathcal{B}_{g}\kappa-6\big) \xi ^2+3
\varphi ^2 (16 \zeta \chi -5)\big)\\\nonumber &+&8 r^9 \xi \varphi
\chi ^3 \big(48 \chi (8 \mathcal{B}_{g}\kappa+14 \zeta \chi +3 \chi
-15) \xi ^4+8 \varphi ^2 (40 \mathcal{B}_{g}\kappa-2 \chi  (25 \chi
\zeta+5\\\nonumber &-&12 \zeta )-15) \xi ^2+3 \varphi ^4 (32 \zeta
\chi -5)\big)+4 r^{15} \xi \chi ^5 \big(16 \chi (10
\mathcal{B}_{g}\kappa+114 \zeta \chi -105)\\\nonumber &\times&
\varphi ^3 \xi ^2+3 \varphi ^2 (24 \mathcal{B}_{g}\kappa+\chi (-54
\chi \zeta +136 \zeta +15)-46)\big)+8 r^{13} \xi \varphi \chi ^4
\big(192 \chi
\\\nonumber &\times& \xi ^4+8  (40 \mathcal{B}_{g}\kappa+222 \zeta \chi
+20 \chi -135)\varphi ^2 \chi \xi ^2+3 \varphi ^4 (8
\mathcal{B}_{g}\kappa+\chi  (2 \zeta (\chi +8)\\\nonumber &-&
5)-5)\big)+4 r^{11} \xi \chi ^3 \big(12 \mathcal{B}_{g}\kappa
\varphi \big(16 \chi ^2 \xi ^4+80 \varphi ^2 \chi \xi ^2+\varphi
^4\big)+\chi \varphi ^5\big(3 (136 \zeta \chi
\\\nonumber &-&25) -8 \xi ^2 (\chi (58 \chi \zeta -264 \zeta
-65)+150) \varphi ^3+\xi ^4 \chi (10 \zeta \chi -13) \varphi
\big)\big)+32 \\\nonumber &\times&r^7 \xi ^3 \varphi \chi ^2 \big(4
\mathcal{B}_{g}\kappa \big(36 \chi \xi ^2+5 \varphi ^2\big)+\chi
\big((2 \chi  (24 \zeta-34 \chi \zeta  +33)-84) \xi ^2+\varphi ^2
\\\nonumber &\times&(224 \zeta \chi
-75)\big)\big)\big)\big)\bigg]\bigg[\kappa (3 s-1) \big(\chi
r^2+1\big)^4 \big(6 r^2 \xi  \varphi ^2 \big(r^2 \chi \big)^{3/2}+8
\xi ^3 \sqrt{r^2 \chi }\\\label{19} &+&r^7 \varphi ^3 \chi ^2+12
r^3 \xi ^2 \varphi  \chi \big)^2\bigg]^{-1},\\
\nonumber p_{t}&=&\bigg[r \big(\chi  r^2+1\big) \big(r^3 \varphi ^3
\big(\chi  \big(\chi  r^4+(2-12 \zeta  \chi ) r^2-8 \zeta
\big)+1\big) \big(r^2 \chi \big)^{3/2}+4 r \xi ^2 \varphi
\\\nonumber &\times&\big(\chi \big(3 \chi  r^4+(6-20 \zeta  \chi )
r^2-8 \zeta \big)+3\big) \sqrt{r^2 \chi }+8 \xi ^3 \big(\chi
\big(\chi r^2-4 \zeta  \chi +2\big) r^2\\\nonumber &+&1\big)+2 r^4
\xi \varphi ^2 \chi \big(\chi  \big(3 \chi r^4+(6-28 \zeta  \chi )
r^2-16 \zeta \big)+3\big)\big) \big(\big[2 \big(\frac{\varphi \chi
r^2}{2 \sqrt{r^2 \chi }}+\frac{1}{2}\\\nonumber &\times& \varphi
\sqrt{r^2 \chi }\big)\big]\big[\xi +\frac{1}{2} r \varphi \sqrt{r^2
\chi }\big]^{-1}\big)^2+\bigg[4 \big(-42 r^8 \zeta \varphi ^3 \chi
^4 \big(r^2 \chi \big)^{3/2}+60 r^8 s \zeta \\\nonumber
&\times&\varphi ^3 \chi ^4 \big(r^2 \chi \big)^{3/2}+112 r^8 \zeta
\xi ^2 \varphi  \chi ^4 \big(r^2 \chi \big)^{3/2}-4 r^2 \varphi ^3
\big(r^2 \chi \big)^{3/2}+10 \big(r^2 \chi \big)^{3/2} \\\nonumber
&\times&r^8 \varphi ^3 \chi ^3-14 r^8 s \varphi ^3 \chi ^3 \big(r^2
\chi \big)^{3/2}+64 r^8 \zeta \varphi ^3 \chi ^3 \big(r^2 \chi
\big)^{3/2}-136 \big(r^2 \chi \big)^{3/2}r^8\\\nonumber &\times& \xi
^2 \varphi \chi ^3 -16 r^2 \xi ^2 \varphi \big(r^2 \chi
\big)^{3/2}+48 s \xi ^2 \varphi \big(r^2 \chi \big)^{3/2}-56 \xi ^2
\varphi \big(r^2 \chi \big)^{3/2}- s \zeta\\\nonumber &\times& \xi
^2 \varphi \chi 2\big(r^2 \chi \big)^{3/2}+64 \zeta  \xi ^2 \varphi
\chi \big(r^2 \chi \big)^{3/2}-4 r^4 \varphi ^3 \big(r^2 \chi
\big)^{5/2}-6 r^2 \varphi ^3 \big(r^2 \chi \big)^{5/2}\\\nonumber
&-& 12 r^2 s \varphi ^3 \big(r^2 \chi \big)^{5/2}+32 \zeta \varphi
^3 \big(r^2 \chi \big)^{5/2}-8 r^2 \xi ^2 \varphi  \big(r^2 \chi
\big)^{5/2}-6 s \xi ^2 \varphi \big(r^2 \chi \big)^{5/2}\\\nonumber
&+&16 \xi ^2 \varphi \big(r^2 \chi \big)^{5/2}-18 r^4 \varphi ^3
\big(r^2 \chi \big)^{7/2}-6 r^{16} \varphi ^3 \chi ^6 \sqrt{r^2 \chi
}+36   \varphi ^3 \chi ^6 \sqrt{r^2 \chi }\zeta \\\nonumber
&\times&r^{14} -22 r^{14} \varphi ^3 \chi ^5 \sqrt{r^2 \chi }+4
r^{12} \varphi ^3 \chi ^5 \sqrt{r^2 \chi }-r^{12} s \varphi ^3 \chi
^5 \sqrt{r^2 \chi }+84 r^{12} \zeta \\\nonumber &\times&\varphi ^3
\chi ^5 \sqrt{r^2 \chi }-40 r^{12} \xi ^2 \varphi  \chi ^5 \sqrt{r^2
\chi }-30 r^{12} \varphi ^3 \chi ^4 \sqrt{r^2 \chi }-120 \varphi ^3
\chi ^4 \sqrt{r^2 \chi }\\\nonumber &\times&r^8 \zeta+384 r^8 s
\zeta \varphi ^3 \chi ^4 \sqrt{r^2 \chi }+32 r^8 \xi ^2 \varphi \chi
^4 \sqrt{r^2 \chi }-12 r^8 s \xi ^2 \varphi  \chi ^4 \sqrt{r^2 \chi
}\\\nonumber &+&176 r^8 \zeta \xi ^2 \varphi \chi ^4 \sqrt{r^2 \chi
}-152 r^6 \zeta  \xi ^2 \varphi  \chi ^4 \sqrt{r^2 \chi }+208 r^6 s
\zeta  \xi ^2 \varphi \chi ^4 \sqrt{r^2 \chi }+4\\\nonumber &\times&
r^8 \varphi ^3 \chi ^3 \sqrt{r^2 \chi }-25 r^8 s \varphi ^3 \chi ^3
\sqrt{r^2 \chi }+16 r^8 \zeta  \varphi ^3 \chi ^3 \sqrt{r^2 \chi }+8
r^6  \chi ^3 \sqrt{r^2 \chi }\zeta\\\nonumber &\times&\varphi ^3+176
r^6 s \zeta \varphi ^3 \chi ^3 \sqrt{r^2 \chi }-168 r^8 \xi ^2
\varphi \chi ^3 \sqrt{r^2 \chi }+72 r^6 \xi ^2 \varphi  \chi ^3
\sqrt{r^2 \chi }-s\\\nonumber &\times&72 r^6  \xi ^2 \varphi \chi ^3
\sqrt{r^2 \chi }+64 r^6 \zeta \xi ^2 \varphi \chi ^3 \sqrt{r^2 \chi
}-416 r^4 \zeta \xi ^2 \varphi  \chi ^3 \sqrt{r^2 \chi }+896 r^4 s
\\\nonumber &\times&\zeta  \xi ^2 \varphi \chi ^3 \sqrt{r^2 \chi }
+48 s \xi ^2 \varphi\sqrt{r^2 \chi }-32 \xi ^2 \varphi \sqrt{r^2
\chi }-28 r^{15} \xi \varphi ^2 \chi ^6+16 r \xi ^3 \chi\\\nonumber
&\times&(1-3 s) +20 r^{13} \xi \varphi ^2 \chi ^5 (6 \zeta \chi
-5)+h \kappa r \big(\chi r^2+1\big)^4 \big(r^3 \varphi ^3 \big(r^2
\chi \big)^{3/2}+r\\\nonumber &\times&12  \xi ^2 \varphi \sqrt{r^2
\chi }+8 \xi ^3+6 r^4 \xi \varphi ^2 \chi \big)+4
\mathcal{B}_{g}\kappa r s \big(\chi r^2+1\big)^4 \big(r^3 \varphi ^3
\big(r^2 \chi \big)^{3/2}\\\nonumber &+&12 r \xi ^2 \varphi
\sqrt{r^2 \chi }+8 \xi ^3+6 r^4 \xi \varphi ^2 \chi \big)+8 r^3 \xi
\chi \big(\chi \xi ^2(-13 s+16 (2 s-1) \zeta  \chi \\\nonumber &-&6)
+3 \varphi ^2 (-16 \zeta \chi s+s+8 \zeta \chi -1)\big)-2 r^{11} \xi
\chi ^4 \big(8 \chi \xi ^2+\varphi ^2 ((3 s-124 \zeta \\\nonumber
&-&10) \chi +66)\big)+4 r^5 \xi \chi ^2 \big(4 (\chi (-4 s+(2 s-1)
\zeta \chi +3)-1) \xi ^2+\varphi ^2 (120\\\nonumber &\times& \zeta
\chi s+3 s+76 \zeta \chi -11)\big)+2 r^7 \xi  \chi ^2 \big(\varphi
^2 (\chi (-27 s+16 \zeta +48 (8 s-3) \zeta \\\nonumber &\times&\chi
+2)-8)-4 \xi ^2 \chi ((s-4 \zeta -2) \chi +6)\big)+4 r^9 \xi \chi ^3
\big(4 \chi  (2 \zeta \chi -3) \xi ^2\\\nonumber &+&\varphi ^2 (\chi
(-39 \chi \zeta +40 \zeta +6 s (9 \zeta \chi
-2)+11)-19)\big)\big)\bigg]\big[3 s-1\big]^{-1}\bigg]\bigg[4 \kappa
r
\\\label{20} &\times&\big(\chi r^2+1\big)^4 \big(2 \xi +r \varphi \sqrt{r^2 \chi
}\big)^3\bigg]^{-1}.
\end{eqnarray}
The expressions for $\rho_{q}$ and $p_{q}$ are derived as follows
\begin{eqnarray}\nonumber
\rho_{q}&=& -\bigg[4 \xi  r^{11} \varphi  \chi ^5 (4
\mathcal{B}_{g}\kappa+36 \zeta \chi +3 h \kappa-42)+4 \xi  r^9
\varphi \chi ^4 (16 \mathcal{B}_{g}\kappa+12 h
\kappa+\chi\\\nonumber &\times& (6 \zeta -3 s+9)-54)+12 \xi r^7
\varphi \chi ^3 (8 \mathcal{B}_{g}\kappa+6 h \kappa+\chi (8 \zeta +6
\zeta (s-2) \chi -s
\\\nonumber &\times&5+6)-10)+4 \xi r^5 \varphi  \chi ^2 (16
\mathcal{B}_{g}\kappa+3 (4 h \kappa+\chi  (4 \zeta ( s-7) \chi -
s+1)-2))\\\nonumber &+&4 \xi r^3 \varphi \chi (4
\mathcal{B}_{g}\kappa+3 h \kappa+3 \chi  (4 \zeta  (5 s-1) \chi -3
s-4))+4 \big(r^2 \chi \big)^{3/2} \big(16
\mathcal{B}_{g}\kappa\\\nonumber &\times& \xi ^2+3 \big(\varphi ^2
(8 \zeta \chi -1)+4 h \kappa \xi ^2+\xi ^2 \chi  (4 \zeta (s-3) \chi
-s+4)\big)\big)+r^2 \big(r^2 \chi \big)^{3/2} \\\nonumber
&\times&\big(4 \mathcal{B}_{g}\kappa \big(24 \xi ^2 \chi +\varphi
^2\big) +3 \big(h \kappa \big(24 \xi ^2 \chi +\varphi ^2\big)-\chi
\big(4 \xi ^2 (\chi  (2 \zeta  s \chi +5 s-5)\\\nonumber
&+&2)+\varphi ^2 (3 (s+3)-8 \zeta (4 s+7) \chi )\big)\big)\big)+r^2
\big(r^2 \chi \big)^{5/2} \big(4 \xi ^2 \chi (16
\mathcal{B}_{g}\kappa+3 (4 h
\\\nonumber &\times&\kappa+\chi  (4 \zeta -s+2)-6)) +\varphi ^2 (16
\mathcal{B}_{g}\kappa+3 (4 h \kappa+\chi  (16 \zeta +12 \zeta (5
s-1) \chi
\\\nonumber &-&7 s-2)-4))\big)+4 \xi ^2 \sqrt{r^2 \chi } (4
\mathcal{B}_{g}\kappa +3 (h \kappa-3 s \chi +\chi ))+r^{12} \varphi
^2 \chi ^5 \sqrt{r^2 \chi }\\\nonumber &\times& (4
\mathcal{B}_{g}\kappa+108 \zeta \chi +3 h \kappa-66)+r^8 \chi ^3
\big(r^2 \chi \big)^{3/2} \big(\varphi ^2 (4 \kappa (4 B+3 h)-3 \chi
(s\\\nonumber &-&84 \zeta -4)-90)-24 \xi ^2 \chi \big)+r^8 \chi ^3
\sqrt{r^2 \chi } \big(4 \xi ^2 \chi  (4 \mathcal{B}_{g}\kappa+3 (4
\zeta \chi +h \kappa-6))\\\nonumber &+&3 \varphi ^2 (8
\mathcal{B}_{g}\kappa+6 h \kappa+\chi (64 \zeta +6 \zeta (s-4) \chi
-5 s+7)-18)\big)-48 \xi r^{13} \varphi \chi ^6\\\nonumber &-&18
r^{14} \varphi ^2 \chi ^6 \sqrt{r^2 \chi }-24 \xi r \varphi \chi
\bigg]\bigg[\kappa (3 s-1) \big(r^2 \chi +1\big)^4 \big(4 \xi r^3
\varphi \chi +4 \xi ^2 \sqrt{r^2 \chi }\\\label{21} &+&r^2 \varphi
^2 \big(r^2 \chi \big)^{3/2}\big) \bigg]^{-1},\\\nonumber
p_{q}&=&\bigg[r \chi \big(144 r^8 \zeta \xi \varphi ^3 \chi ^4
\big(r^2 \chi \big)^{3/2}-48 r^8 s \zeta \xi \varphi ^3 \chi ^4
\big(r^2 \chi \big)^{3/2}- r^8   \xi ^3 \varphi \chi ^4 \big(r^2
\chi \big)^{3/2}\\\nonumber &\times&256\zeta+24 r^2 \xi \varphi ^3
\big(r^2 \chi \big)^{3/2}-52 r^8 \xi  \varphi ^3 \chi ^3 \big(r^2
\chi \big)^{3/2}+40 r^8 s \xi  \varphi ^3 \chi ^3 \big(r^2 \chi
\big)^{3/2}\\\nonumber &-&288 r^8 \zeta  \xi  \varphi ^3 \chi ^3
\big(r^2 \chi \big)^{3/2}+320 r^8 \xi ^3 \varphi  \chi ^3 \big(r^2
\chi \big)^{3/2}+32 r^2 \xi ^3 \varphi  \big(r^2 \chi \big)^{3/2}+96
\\\nonumber &\times&s \xi ^3 \varphi  \big(r^2 \chi \big)^{3/2}+48
\xi ^3 \varphi \big(r^2 \chi \big)^{3/2}-320 s \zeta \xi ^3 \varphi
\chi \big(r^2 \chi \big)^{3/2}+  \xi ^3 \varphi \chi \big(r^2 \chi
\big)^{3/2}\\\nonumber &\times&64\zeta+24 r^4 \xi \varphi ^3
\big(r^2 \chi \big)^{5/2}+52 r^2 \xi  \varphi ^3 \big(r^2 \chi
\big)^{5/2}+24 r^2 s \xi \varphi ^3 \big(r^2 \chi
\big)^{5/2}-12\\\nonumber &\times& \zeta  \xi \varphi ^3 \big(r^2
\chi \big)^{5/2}+192 r^2 \xi ^3 \varphi \big(r^2 \chi \big)^{5/2}+24
s \xi ^3 \varphi \big(r^2 \chi \big)^{5/2}-80 \xi ^3 \varphi
\big(r^2 \chi \big)^{5/2}\\\nonumber &+&112 r^4 \xi  \varphi ^3
\big(r^2 \chi \big)^{7/2}+40 r^{16} \xi \varphi ^3 \chi ^6 \sqrt{r^2
\chi }-192 r^{14} \zeta  \xi \varphi ^3 \chi ^6 \sqrt{r^2 \chi
}+144\xi\\\nonumber &\times&r^{14}  \varphi ^3 \chi ^5 \sqrt{r^2
\chi }-28 r^{12} \xi \varphi^3 \chi ^5 \sqrt{r^2 \chi }+8 r^{12} s
\xi \varphi ^3 \chi ^5 \sqrt{r^2 \chi }-416 r^{12} \zeta  \xi
\varphi ^3 \\\nonumber &\times&\chi ^5 \sqrt{r^2 \chi }+96 r^{12}
\xi ^3 \varphi  \chi ^5 \sqrt{r^2 \chi }+192 r^{12} \xi \varphi ^3
\chi ^4 \sqrt{r^2 \chi }+160 r^8 \zeta \xi  \varphi ^3 \chi
^4\\\nonumber &\times& \sqrt{r^2 \chi }-432 r^8 s \zeta  \xi \varphi
^3 \chi ^4 \sqrt{r^2 \chi }-80 r^8 \xi ^3 \varphi \chi ^4 \sqrt{r^2
\chi }+32 r^8 s \xi ^3 \varphi \chi ^4 \sqrt{r^2 \chi }\\\nonumber
&-&384 r^8 \zeta  \xi ^3 \varphi \chi ^4 \sqrt{r^2 \chi }+192 r^6
\zeta  \xi ^3 \varphi \chi ^4 \sqrt{r^2 \chi }-64 r^6 s \zeta \xi ^3
\varphi  \chi ^4 \sqrt{r^2 \chi }+4 r^8\\\nonumber &\times& \xi
\varphi ^3 \chi ^3 \sqrt{r^2 \chi }+56 r^8 s \xi \varphi ^3 \chi ^3
\sqrt{r^2 \chi }-64 r^8 \zeta  \xi \varphi ^3 \chi ^3 \sqrt{r^2 \chi
}-208 r^6 \zeta \xi \varphi ^3 \chi ^3 \\\nonumber &\times&\sqrt{r^2
\chi }-208 r^6 s \zeta  \xi \varphi ^3 \chi ^3 \sqrt{r^2 \chi }+384
r^8 \xi ^3 \varphi  \chi ^3 \sqrt{r^2 \chi }-6 r^6 \xi ^3 \varphi
\chi ^3 \sqrt{r^2 \chi }\\\nonumber &+&60 r^6 s \xi ^3 \varphi \chi
^3 \sqrt{r^2 \chi }-128 r^6 \zeta \xi ^3 \varphi \chi ^3 \sqrt{r^2
\chi }+64 r^4 \zeta  \xi ^3 \varphi \chi ^3 \sqrt{r^2 \chi }-1088
\\\nonumber &\times&r^4 s \zeta \xi ^3 \varphi \chi ^3 \sqrt{r^2 \chi }
+32 \xi ^3\varphi \sqrt{r^2 \chi }+6 r^{19} \varphi ^4 \chi ^7+16 r
(3 s-1) \xi ^4 \chi + r^{17}\\\nonumber &\times&\chi ^6 (11-18 \zeta
\chi )-h \kappa r \big(\chi r^2+1\big)^4 \big(8 r^3 \xi \varphi ^3
\big(r^2 \chi \big)^{3/2}+ r \xi ^3 \varphi \sqrt{r^2 \chi }+ \xi
^4\\\nonumber &+&r^8 \varphi ^4 \chi ^2+24 r^4 \xi ^2 \varphi ^2
\chi \big)-4 \mathcal{B}_{g}\kappa r s \big(\chi r^2+1\big)^4 \big(8
r^3 \xi \varphi ^3 \big(r^2 \chi \big)^{3/2}+32 r \xi ^3
\\\nonumber &\times&\sqrt{r^2 \chi }+16 \xi ^4+r^8 \varphi ^4 \chi
^2+24 r^4 \xi ^2 \varphi ^2 \chi \big)-16 r^3 \xi ^2 \chi \big(\chi
(-7 s+4 (5 s-3)\\\nonumber &\times& \zeta \chi +4) \xi ^2+\varphi ^2
(8 \zeta \chi -3)\big)+r^{15} \varphi ^2 \chi ^5 \big( \chi \xi
^2+\varphi ^2 ((s-4 \zeta -4) \chi +30)\big)\\\nonumber &+&r^9 \chi
^3 \big(32 \chi  (3-2 \zeta \chi ) \xi ^4- \varphi ^2 \big(2 (7
s-18) \zeta \chi ^2+(6-5 s+16 \zeta) 3\chi -30\big)\\\nonumber
&\times& \xi ^2+\varphi ^4 (3 (s+3)-8 (4 s+7) \zeta \chi )\big)+8
r^5 \xi ^2 \chi ^2 \big(2 (\chi ( \zeta  \chi s+5 s-5)+2)\\\nonumber
&\times& \xi ^2+\varphi ^2 (-56 \zeta  \chi s+9 s-20 \zeta  \chi
+12)\big)+4 r^7 \chi ^2 \big(4 \chi ((s-4 \zeta -2) \chi
+6)\\\nonumber &\times& \xi ^4+2 \varphi ^2 (\chi (21 s-8 \zeta +
(1-2 s) \zeta  \chi -3)+6) \xi ^2+\varphi ^4 (1-8 \zeta \chi
)\big)+r^{13} \\\nonumber &\times&\varphi ^2 \chi ^4 \big(16 \chi
(21-22 \zeta \chi ) \xi ^2+\varphi ^2 (\chi  (5 s-64 \zeta -6 (s-4)
\zeta \chi -7)+18)\big)\\\nonumber &+&r^{11} \chi ^3 \big(32 \chi ^2
\xi ^4+24 \varphi ^2 \chi  ((s-28 \zeta -3) \chi +18) \xi ^2+\varphi
^4 (\chi (-60 \zeta \chi  s+7 s\\\nonumber &-&16 \zeta +12 \zeta
\chi +2)+4)\big)\big)\bigg]\bigg[\kappa (3 s-1) \big(\chi
r^2+1\big)^4 \big(r^2 \varphi ^2 \big(r^2 \chi \big)^{3/2}+4 \xi
^2\\\label{22} &\times& \sqrt{r^2 \chi }+4 r^3 \xi  \varphi  \chi
\big)^2\bigg]^{-1}.
\end{eqnarray}

\section*{Appendix C: Boundary Condition at the Stellar Surface}
\renewcommand{\theequation}{C\arabic{equation}}
\setcounter{equation}{0}

To maintain physical continuity, the radial pressure must vanish at
the boundary of the star, i.e., $p_r(R) = 0$. By applying the linear
equation of state $p_r = s\rho - h$ and substituting the expressions
for $B_1 $ and $ B_2$ into the resulting field equations, we derive
the following explicit expression for the constant $h$
\begin{eqnarray}\nonumber
h&=&-\bigg[s \big(64 \xi ^4 \big(16 \mathcal{B}_{g}\kappa \xi
^2+\chi (16 \zeta \chi -5) \xi ^2+3 \varphi ^2 (5-8 \zeta  \chi
)\big) \big(r^2 \chi \big)^{3/2}-32\\\nonumber&\times& r^2 \xi ^4
\big(\chi \big(4 (\chi (\zeta \chi -5)+3) \xi ^2+3 \varphi ^2 (25-44
\zeta \chi )\big)- \big(8 \chi  \xi ^2+5 \varphi ^2\big)6 B
\kappa\big)
\\\nonumber&\times&\big(r^2 \chi \big)^{3/2}+16 r^2 \xi ^2 \big(4 \chi  (16
\mathcal{B}_{g}\kappa+12 \zeta \chi +5 \chi -18) \xi ^4+3 \varphi ^2
(80 \mathcal{B}_{g}\kappa+\chi \\\nonumber&\times& (8 \zeta  (2-15
\chi )-5)-20) \xi ^2+6 \varphi ^4 (24 \zeta  \chi -5)\big) \big(r^2
\chi \big)^{5/2}+16 r^4 \xi ^2 \big(\mathcal{B}_{g}\kappa
\\\nonumber&\times&\big(16 \chi ^2 \xi ^4+360 \varphi ^2 \chi \xi
^2+15 \varphi ^4\big)+\chi \big(24 \chi (2 \zeta  \chi -3) \xi ^4-6
(\chi  (53 \chi \zeta -64\\\nonumber&\times& \zeta -30)+55)\varphi
^2 \xi ^2+\varphi^4 (316 \zeta  \chi -75)\big)\big) \big(r^2 \chi
\big)^{5/2}-4 r^4 \big(96 \chi ^2 \xi ^6-12
\\\nonumber&\times&\varphi ^2 \chi (80 B \kappa+268 \zeta  \chi +35 \chi
-210) \xi ^4+\varphi ^4 (-240 \mathcal{B}_{g}\kappa+\chi (32 \zeta
(5 \chi -9)\\\nonumber&+&105)+120) \xi ^2+3 \varphi ^6 (1-8 \zeta
\chi )\big) \big(r^2 \chi \big)^{7/2}-18 r^{22} \varphi ^6 \chi ^8
\sqrt{r^2 \chi }+256 \mathcal{B}_{g}\kappa\\\nonumber&\times& \xi ^6
\sqrt{r^2 \chi }+2 r^{20} \varphi ^6 \chi ^7 (2
\mathcal{B}_{g}\kappa+54 \zeta \chi -33) \sqrt{r^2 \chi }+r^{18}
\varphi ^4 \chi ^6 \big(\varphi ^2 (16 \mathcal{B}_{g}
\kappa\\\nonumber&+&252 \zeta \chi +11 \chi -90)-840 \xi ^2 \chi
\big) \sqrt{r^2 \chi }+2 r^{16} \varphi ^4 \chi ^5 \big(12 \chi  (10
\mathcal{B}_{g}\kappa+158\\\nonumber&\times& \zeta \chi -125) \xi
^2+\varphi ^2 (12 \mathcal{B}_{g}\kappa+\chi (-33 \chi \zeta +96
\zeta +8)-27)\big) \sqrt{r^2 \chi }+r^{14}\\\nonumber&\times&
\varphi ^2 \chi ^4 \big(-2400 \chi ^2 \xi ^4+12 \varphi ^2 \chi  (80
\mathcal{B}_{g}\kappa+668 \zeta \chi +45 \chi -330) \xi ^2+\varphi
^4
\\\nonumber&\times&(16 \mathcal{B}_{g}\kappa+\chi (24 \zeta (\chi
+2)-13)-12)\big) \sqrt{r^2 \chi }+2 r^{12} \varphi ^2 \chi ^3 \big(2
\mathcal{B}_{g}\kappa \big(24 \chi ^2 \\\nonumber&\times&\xi ^4+360
\varphi ^2 \chi \xi ^2+\varphi ^4\big)+\chi \big(48 \chi  (78 \zeta
\chi -85) \xi ^4+4 \varphi ^2 (\chi  (319 \chi  \zeta
+\zeta\\\nonumber&\times& 672 +105)-285) \xi ^2+5 \varphi ^4 (20
\zeta \chi -3)\big)\big) \sqrt{r^2 \chi }-192 r^{21} \xi \varphi ^5
\chi ^8-3\\\nonumber&\times&r \xi ^5 \varphi \chi +24 r^{19} \xi
\varphi ^5 \chi ^7 (2 \mathcal{B}_{g}\kappa+42 \zeta \chi -29)+64
r^3 \xi ^5 \varphi \chi (12
\mathcal{B}_{g}\kappa+\chi\\\nonumber&\times& (8 \zeta \chi -15))+24
r^{17} \xi \varphi ^3 \chi ^6 \big(\varphi ^2 (8
\mathcal{B}_{g}\kappa+94 \zeta \chi +5 \chi -39)-80 \xi ^2 \chi
\big)\\\nonumber&+&64 r^5 \xi ^3 \varphi \chi ^2 \big(\big(68 \zeta
\chi ^2+6 \chi +48 \mathcal{B}_{g}\kappa-6\big) \xi ^2+3 \varphi ^2
(16 \zeta  \chi -5)\big)+8 r^9\\\nonumber&\times& \xi \varphi  \chi
^3 \big(48 \chi (8 \mathcal{B}_{g}\kappa+14 \zeta \chi +3 \chi -15)
\xi ^4+8 \varphi ^2 (40 \mathcal{B}_{g}\kappa-2 \chi (225 \chi \zeta
\\\nonumber&-&12 \zeta +5)-15) \xi ^2+3 \varphi ^4 (32 \zeta  \chi
-5)\big)+4 r^{15} \xi \varphi ^3 \chi ^5 \big(16 \chi (10
\mathcal{B}_{g}\kappa+ \zeta\\\nonumber&\times& 114 \chi -105) \xi
^2+3 \varphi ^2 (24 \mathcal{B}_{g}\kappa+\chi (-54 \chi \zeta +136
\zeta +15)-46)\big)+8
\\\nonumber&\times&r^{13} \xi \varphi \chi ^4 \big(-192 \chi ^2
\xi ^4+8 \varphi ^2\chi (40 \mathcal{B}_{g}\kappa+222 \zeta \chi +20
\chi -135) \xi ^2+3\\\nonumber&\times&\varphi ^4 (8
\mathcal{B}_{g}\kappa+\chi (2 \zeta (\chi +8)-5)-5)\big)+4 r^{11}
\xi \chi ^3 \big(12 \mathcal{B}_{g}\kappa \varphi \big(16 \chi ^2
\xi ^4+8\\\nonumber&\times& \varphi ^2 \chi \xi ^2+\varphi
^4\big)+\chi \big(3 (136 \zeta \chi -25) \varphi ^5-8 \xi ^2 (\chi
(158 \chi \zeta -264 \zeta -65)\\\nonumber&+&150) \varphi ^3+96 \xi
^4 \chi (10 \zeta \chi -13) \varphi \big)\big)+32 r^7 \xi ^3 \varphi
\chi ^2 \big(4 \mathcal{B}_{g}\kappa \big(36 \chi \xi ^2+5 \varphi
^2\big)\\\nonumber&+&\chi \big((2 \chi (34 \chi \zeta +24 \zeta
+33)-4) \xi ^2+\varphi ^2 (24 \zeta \chi
-75)\big)\big)\big)\bigg]\bigg[\kappa \big(\chi r^2+1\big)^4
\\\nonumber&\times&\big(24 r^2 \xi ^4 \varphi ^2 \big(r^2 \chi
\big)^{3/2}+60 r^4 \xi ^2 \varphi ^4 \big(r^2 \chi \big)^{5/2}+64
\xi ^6 \sqrt{r^2 \chi }+ \chi ^3 \sqrt{r^2 \chi
}r^{12}\\\label{aa}&\times& \varphi ^6+12 r^{11} \xi \varphi ^5 \chi
^3+160 r^7 \xi ^3 \varphi ^3 \chi ^2+192 r^3 \xi ^5 \varphi \chi
\big)\bigg]^{-1}.
\end{eqnarray}

\section*{Appendix D: Calculation of the Bounds of the Coupling Parameters}
\renewcommand{\theequation}{B\arabic{equation}}
\setcounter{equation}{0}

We determine the following limit for $\zeta$
\begin{eqnarray}\nonumber
\zeta&>&\big[\big(\chi  r^2+1\big)^2 \big(\big[18 r^8 \varphi ^2
\chi ^3 \big(r^2 \chi \big)^{3/2}+r^2 \big(2 \big(3 \varphi ^2-2 (4
\mathcal{B}_{g} \kappa+3 h \kappa-6) \xi ^2\big) \chi
\\\nonumber&-&(4 \mathcal{B}_{g}+3 h) \kappa \varphi ^2\big) \big(r^2
\chi \big)^{3/2}-4 \big(\xi ^2 (8 \mathcal{B}_{g} \kappa+6 h
\kappa+5 \chi )-3 \varphi ^2\big) \big(r^2 \chi
\big)^{3/2}\\\nonumber&+&r^2 \big(24 \xi ^2 \chi -\varphi ^2 (8 B
\kappa+6 h \kappa+11 \chi -12)\big) \big(r^2 \chi \big)^{5/2}+(-4
\mathcal{B}_{g} \kappa-3 h \kappa+30)\\\nonumber&\times& r^8 \varphi
^2 \chi ^3 \sqrt{r^2 \chi }-4 (4 \mathcal{B}+3 h) \kappa \xi ^2
\sqrt{r^2 \chi }+48 r^9 \xi \varphi \chi ^4-4 (4 \mathcal{B}_{g}
\kappa+3 h \kappa-18)\\\nonumber&\times&r^7 \xi \varphi  \chi ^3+24
r \xi \varphi \chi -4 r^3 \xi \varphi (4 \mathcal{B}_{g} \kappa+3 h
\kappa-3 \chi ) \chi -8 r^5 \xi \varphi \chi ^2 (4 \mathcal{B}_{g}
\kappa+3 h \kappa\\\nonumber&+&4 \chi -3)\big]\big[54 r^8 \varphi ^2
\chi ^3 \big(r^2 \chi \big)^{3/2}-4 \big(\xi ^2 \chi -25 \varphi
^2\big) \big(r^2 \chi \big)^{3/2}+12 r^2 \big(2 \chi \xi ^2+\varphi
^2
\\\nonumber&\times&(\chi +2)\big) \big(r^2 \chi \big)^{3/2}+3 r^2
\big(8 \chi \xi ^2+\varphi ^2 (32-11 \chi )\big) \big(r^2 \chi
\big)^{5/2}+ r^8 \varphi ^2 \chi ^3 \sqrt{r^2 \chi }\\\nonumber&+&16
\big(3 \varphi ^2-2 \xi ^2 \chi \big) \sqrt{r^2 \chi }+72 r^9 \xi
\varphi \chi ^4+120 r^7 \xi \varphi \chi ^3-72 r^3 \xi  \varphi \chi
^2+16 r \xi  \varphi \chi\\\nonumber&+&12 r^5 \xi \varphi \chi ^2
(4-5 \chi )\big]^{-1}-\big[s \big(192 \xi ^5 \varphi \big(r^2 \chi
\big)^{3/2}-192 r^2 \xi ^3 \varphi \big(\xi ^2 (6 \chi -2)-5
\\\nonumber&\times&\varphi ^2\big) \big(r^2 \chi \big)^{3/2}+480 r^2 \xi
^3 \varphi \big(4 \xi ^2+\varphi ^2\big) \big(r^2 \chi \big)^{5/2}+8
r^4 \xi \varphi \big(192 \chi \xi ^4+40 \varphi ^2 (3\\\nonumber&-&4
\chi ) \xi ^2+15 \varphi ^4\big) \big(r^2 \chi \big)^{5/2}+60 r^4
\xi \varphi ^3 \big(48 \xi ^2+\varphi ^2\big) \big(r^2 \chi
\big)^{7/2}+192 r^{16} \xi \varphi ^5 \chi ^5\\\nonumber&\times&
\sqrt{r^2 \chi }+312 r^{14} \xi \varphi ^5 \chi ^4 \sqrt{r^2 \chi
}+384 \xi ^5 \varphi \sqrt{r^2 \chi }+120 r^{12} \xi \varphi ^3 \chi
^3 \big(16 \xi ^2 \chi -\varphi ^2\\\nonumber&\times& (\chi -1)\big)
\sqrt{r^2 \chi }+18 r^{19} \varphi ^6 \chi ^6+30 r^{17} \varphi ^6
\chi ^5+6 r^{13} \varphi ^4 \big(220 \xi ^2+\varphi ^2\big) \chi
^4+240 \\\nonumber&\times&r^9 \xi ^2 \varphi ^2 \big(14 \xi
^2+\varphi ^2\big) \chi ^3+96 r^5 \xi ^4 \big(4 \xi ^2+5 \varphi
^2\big) \chi ^2-320 r^3 \xi ^4 \chi \big(\xi ^2 \chi -3 \varphi
^2\big)\\\nonumber&+&r^{15} \varphi ^4 \chi ^4 \big(840 \chi \xi
^2+\varphi ^2\big)+48 r^7 \xi ^2 \chi ^2 \big(8 \chi \xi ^4+5
\varphi ^2 (4-7 \chi ) \xi ^2+10 \varphi
^4\big)+2\\\nonumber&\times& r^{11} \varphi ^2 \chi ^3 \big(200 \chi
\xi ^4+5 \varphi ^2 (8-9 \chi ) \xi ^2+\varphi ^4\big)-4
\mathcal{B}_{g} \kappa r \big(\chi r^2+1\big)^2 \big(160 r^3 \xi ^3
\varphi ^3 \\\nonumber&\times&\big(r^2 \chi \big)^{3/2}+12 r^5 \xi
\varphi ^5 \big(r^2 \chi \big)^{5/2}+192 r \xi ^5 \varphi \sqrt{r^2
\chi }+64 \xi ^6+r^{12} \varphi ^6 \chi ^3+60 r^8
\\\nonumber&\times&\xi ^2 \varphi ^4 \chi ^2+240 r^4 \xi ^4 \varphi
^2 \chi \big)\big)-h \kappa r \big(\chi r^2+1\big)^2 \big(160 r^3
\xi ^3 \varphi ^3 \big(r^2 \chi \big)^{3/2}+12 r^5
\\\nonumber&\times&\xi \varphi ^5 \big(r^2 \chi \big)^{5/2}+192 r
\xi ^5 \varphi \sqrt{r^2 \chi }+60 r^8 \xi ^2 \varphi ^4 \chi ^2+240
r^4 \xi ^4 \varphi ^2 \chi \big)\big]=\Psi,
\end{eqnarray}
where $\Psi$ represents the lower bound obtained from the trace
energy condition. Additionally, since the transverse sound velocity
inside the star gives us another limit for $\zeta$
\begin{eqnarray}\nonumber
\zeta&<&1+\bigg[r^2 \big(2 \big(3 \varphi ^2-2 (4 \mathcal{B}_{g}
\kappa+18 r^8 \varphi ^2 \chi ^3 \big(r^2 \chi \big)^{3/2}+3 h
\kappa-6) \xi ^2\big) \chi +30\mathcal{B}_{g} \kappa\\\nonumber&-&(4
\mathcal{B}_{g}+3 h) \kappa \varphi ^2\big) \big(r^2 \chi
\big)^{3/2}-4 \big(\xi ^2 (8 \mathcal{B}_{g} \kappa+6 h \kappa+5
\chi )-3 \varphi ^2\big) \big(r^2 \chi \big)^{3/2}\\\nonumber&+&r^2
\big(24 \xi ^2 \chi -\varphi ^2 (8 B \kappa+6 h \kappa+11 \chi
-12)\big) \big(r^2 \chi \big)^{5/2}+(-4 \mathcal{B}_{g} \kappa-3 h
\kappa)\\\nonumber&\times& r^8 \varphi ^2 \chi ^3 \sqrt{r^2 \chi }-4
(4 \mathcal{B}+3 h) \kappa \xi ^2 \sqrt{r^2 \chi }+48 r^9 \xi
\varphi \chi ^4-4 (4 \mathcal{B}_{g} \kappa+3 h
\kappa-18)\\\nonumber&\times&r^7 \xi \varphi  \chi ^3+24 r \xi
\varphi \chi -4 r^3 \xi \varphi (4 \mathcal{B}_{g} \kappa+3 h
\kappa-3 \chi ) \chi -8 r^5 \xi \varphi \chi ^2 (4 \mathcal{B}_{g}
\kappa+3 h \kappa\\\nonumber&\times& r^{11} \varphi ^2 \chi ^3
\big(200 \chi \xi ^4+5 \varphi ^2 (8-9 \chi ) \xi ^2+\varphi
^4\big)-4 \mathcal{B}_{g} \kappa r \big(\chi r^2+1\big)^2 \big(160
r^3 \xi ^3 \varphi ^3
\\\nonumber&\times&\big(r^2 \chi \big)^{3/2}+12 r^5 \xi \varphi ^5
\big(r^2 \chi \big)^{5/2}+192 r \xi ^5 \varphi \sqrt{r^2 \chi }+64
\xi ^6+r^{12} \varphi ^6 \chi ^3+60 r^8
\\\nonumber&\times&\xi ^2 \varphi ^4 \chi ^2+240 r^4 \xi ^4 \varphi
^2 \chi \big)\big)-h \kappa r \big(\chi r^2+1\big)^2-4
\mathcal{B}_{g} \kappa r \big(\chi r^2+1\big)^2 24 \xi ^2
\\\nonumber&\times&r^9 \xi ^2 \varphi ^2 \big(14 \xi ^2+\varphi
^2\big) \chi ^3+96 r^5 \xi ^4 \big(4 \xi ^2+5 \varphi ^2\big) \chi
^2-320 r^3 \xi ^4 \chi \big(\xi ^2 \chi -3 \varphi
^2\big)\\\nonumber&+&r^{15} \varphi ^4 \chi ^4 \big(840 \chi \xi
^2+\varphi ^2\big)+48 r^7 \xi ^2 \chi ^2 \big(8 \chi \xi ^4+5
\varphi ^2 (4-7 \chi ) \xi ^2+10 \varphi
^4\big)+2\\\nonumber&\times& r^{11} \varphi ^2 \chi ^3 \big(200 \chi
\xi ^4+5 \varphi ^2 (8-9 \chi ) \xi ^2+\varphi ^4\big)-4
\mathcal{B}_{g} \kappa r \big(\chi r^2+1\big)^2 \big(160 r^3 \xi ^3
\varphi ^3 \\\nonumber&\times&\big(r^2 \chi \big)^{3/2}+12 r^5 \xi
\varphi ^5 \big(r^2 \chi \big)^{5/2}+192 r \xi ^5 \varphi \sqrt{r^2
\chi }+64 \xi ^6+r^{12} \varphi ^6 \chi ^3+60 r^8\\\nonumber&+&4
\chi -3)\bigg]\bigg[54 r^8 \varphi ^2 \chi ^3 \big(r^2 \chi
\big)^{3/2}-4 \big(\xi ^2 \chi -25 \varphi ^2\big) \big(r^2 \chi
\big)^{3/2}+12 r^2 \big(2 \chi \xi ^2+\varphi ^2
\\\nonumber&\times&(\chi +2)\big) \big(r^2 \chi \big)^{3/2}+3 r^2
\big(8 \chi \xi ^2+\varphi ^2 (32-11 \chi )\big) \big(r^2 \chi
\big)^{5/2}+ r^8 \varphi ^2 \chi ^3 \sqrt{r^2 \chi }\\\nonumber&+&16
\big(3 \varphi ^2-2 \xi ^2 \chi \big) \sqrt{r^2 \chi }+72 r^9 \xi
\varphi \chi ^4+120 r^7 \xi \varphi \chi ^3-72 r^3 \xi  \varphi \chi
^2+16 r \xi  \varphi \chi\\\nonumber&+&12 r^5 \xi \varphi \chi ^2
(4-5 \chi )\bigg]^{-1}=1+\Phi.
\end{eqnarray}
Here, $1 +\Phi$ represents the upper bound derived from the sound
velocity constraint.\\\\
\textbf{Data Availability Statement:} No data was used for the
research described in this paper.


\begin{thebibliography}{55}

\bibitem{1} Riess, A.G. et al.: Astron. J. \textbf{116}(1998)1009;
Eisenstein, D.J. et al.: Astrophys. J. \textbf{633}(2005)560;
Komatsu, E. et al.: Astrophys. J. Suppl. \textbf{180}(2009)330.

\bibitem{2} Beutler, F. et al.: Mon. Not. Roy. Astron. Soc.
\textbf{416}(2011)3017; Betoule, M. et al.: Astron. Astrophys.
\textbf{568}(2014)32; Ade, P. et al.: Astron. Astrophys.
\textbf{594}(2016)28; Aghanim, N. et al.: Astron. Astrophys.
\textbf{594}(2016)99.

\bibitem{5} Cartan, $\acute{E}$.: C.R. and Acad. Sci. Paris
\textbf{174}(1922)593.

\bibitem{6} Weitzenb$\ddot{o}$ck, R.: \emph{Invariantentheorie}
(Noordhoff, Groningen, 1923); M$\ddot{o}$ller, C. and Dan, K.: Mat.
Fys. Skr. \textbf{1}(1961)10; Pellegrini, C. and Plebanski, J.: K.
Dan. Vidensk. Selsk., Mat. Fys. Skr. \textbf{2}(1963)4; Hayashi, K.
and Shirafuji, T.: Phys. Rev. D \textbf{19}(1979)3524.

\bibitem{7} Nester, J.M. and  Yo, H.-J.: Chin. J. Phys. \textbf{37}(1999)113;
Adak, M., Kalay, M. and Sert, O.: Int. J. Mod. Phys. D
\textbf{15}(2006)619.

\bibitem{8} Jim$\acute{e}$nez, J.B., Heisenberg, L. and Koivisto, T.:
Phys. Rev. D \textbf{98}(2018)044048; Delhom-Latorre, A., Olmo, G.J.
and Ronco, M.: Phys. Lett. B \textbf{780}(2018)294; Harko, T. et
al.: Phys. Rev. D \textbf{98}(2018)084043.

\bibitem{9} Jim$\acute{e}$nez, J.B., Heisenberg, L. and Koivisto, T.: Universe
\textbf{5}(2019)173; Jim$\acute{e}$nez, J.B. et al.: Phys. Rev. D
\textbf{101}(2020)103507.

\bibitem{66b} Abbas, G. and  Nazar, H.: Ann. Phys. \textbf{424}(2021)168336.

\bibitem{66c} Frolov, A.V.: Phys. Rev. Lett. \textbf{101}(2008)061103;
Reverberi, L.: Phys. Rev. D \textbf{87}(2013)084005.

\bibitem{66d} Saha, P. and Debnath, U.: Adv. High Energy Phys.
\textbf{2018}(2018)3901790.

\bibitem{66e} Camera, S., Cardone, V.F. and Radicella, N.: Phys. Rev. D
\textbf{89}(2014)083520.

\bibitem{13} Anagnostopoulos, F.K. et al.: Phys. Lett. B
\textbf{822}(2021)136634.

\bibitem{14} Frusciante, N.: Phys. Rev. D \textbf{103}(2021)044021.

\bibitem{16} Mandal, S., Parida, A. and Sahoo, P.K.: Universe \textbf{8}(2022)240.

\bibitem{18} Wang, W., Chen, H. and Katsuragawa, T.: Phys. Rev. D \textbf{105}(2022)
024060.

\bibitem{19} D'Ambrosio, F. et al.: Phys. Rev. D \textbf{105}(2022)024042.

\bibitem{20} Gadbail, G.N., Mandal, S. and Sahoo, P.K.: Phys. Lett. B
\textbf{835}(2022)137509.

\bibitem{21} De, A. and Loo, T.H.: Class. Quantum Grav. \textbf{40}(2023)115007;
Khyllep, W. et al.: Phys. Rev. D \textbf{107}(2023)044022; Koussour,
M. et al.: Prog. Theor. Exp. Phys. \textbf{2023} (2023)113E01;
Sharif, M. and Ajmal, M.: Chin. J. Phys. \textbf{88}(2024)706; Phys.
Scr. \textbf{99}(2024)085039; Phys. Dark Universe
\textbf{46}(2024)101572; Astropart. Phys. \textbf{165}(2025)103054
Sharif, M., Gul, M.Z. and Fatima, N.: New Astron.
\textbf{109}(2024)102211; Gadbail, G.N., Mandal, S. and Sahoo, P.
K.: Astrophys. J. \textbf{972}(2024)174.

\bibitem{21a} Zhang, C. and Ren, J.: Phys. Rev. D \textbf{108}(2023)063012.

\bibitem{22} Mishra, H. et al.: Int. J. Mod. Phys. E \textbf{2}(1993)547.

\bibitem{24} Khadkikar, S.B., Mishra, A. and Mishra, H.: Mod. Phys. Lett.
A \textbf{10}(1995)2651.

\bibitem{25} Maheswari, V.U. et al.: Nucl. Phys. A \textbf{615}(1997)516.

\bibitem{26} Schertler, K. et al.: Nucl. Phys. A \textbf{637}(1998)451.

\bibitem{28} Blackman, E.G., Frank, A. and Welch, C.:
Astrophys. J. \textbf{546}(2001)288.

\bibitem{29} Gupta, V. K., Tuli, V. and Goyal, A.:
Astrophys. J. \textbf{579}(2002)374.

\bibitem{30} Grigorian, H., Blaschke, D. and Aguilera, D.N.:
Phys. Rev. C-Nucl. Phys. \textbf{69}(2004)065802.

\bibitem{30a} Alford, M. et al.: Astrophys. J. \textbf{629}(2005)969.

\bibitem{30b} Nicotra, O.E. et al.: Phys. Rev. D \textbf{74}(2006)123001.

\bibitem{31} Hussain, H. et al.: J. Polym. Sci. Polym. Chem.
\textbf{46}(2008)7287.

\bibitem{34} Dexheimer, V., Negreiros, R. and Schramm, S.: Eur. Phys.
J. A \textbf{48}(2012)1.

\bibitem{34a} Alford, M.G., Han, S. and Prakash, M.: Phys. Rev. D
\textbf{88}(2013)083013.

\bibitem{35} Bhar, P.: Astrophys. Space Sci. \textbf{357}(2015)1.

\bibitem{36} Burgio, G.F. and Zappala, D.: Eur. Phys. J. A
\textbf{52}(2016)60.

\bibitem{36a} Kaltenborn, M.A.R., Bastian, N.U.F. and  Blaschke, D.B.: Phys. Rev.
D \textbf{96}(2017)056024.

\bibitem{37} Nandi, R. and Char, P.: Astrophys. J. \textbf{857}(2018)12.

\bibitem{39} Khanmohamadi, S., Moshfegh, H.R. and Tehrani, S.A.:
Phys. Rev. D \textbf{101}(2020)023004.

\bibitem{42} Laskos-Patkos, P., Koliogiannis, P.S. and Moustakidis, C.C.:
Phys. Rev. C \textbf{109}(2024)063017; Mariani, M. et al.: Phys.
Rev. C \textbf{110}(2024)043026; Rigtering, C. et al.: J. Bus. Res.
\textbf{176}(2024)114596; Li, J.J., Sedrakian, A. and Alford, M.:
Astrophys. J. \textbf{967}(2024)116.

\bibitem{40} Bhar, P. et al.: Eur. Phys. J. C \textbf{83}(2023)737.

\bibitem{40a} Rej, P.: Chin. J. Phys. \textbf{89}(2024)174.

\bibitem{43} Finch, M.R. and Skea, J.E.F.: Class. Quantum Grav. \textbf{6}(1989)467.

\bibitem{44} Hansraj, S. et al.: Int. J. Mod. Phys. D \textbf{15}(2006)1311;
Banerjee, A. et al.: Gen. Relativ. Gravit. \textbf{45}(2013)717.

\bibitem{45} Sharma, R. and Ratanpal, B.S.: Int. J. Mod. Phys. D
\textbf{22}(2013)1350074.

\bibitem{46} Sharma R. and Das S.: J. Gravit. \textbf{2013}(2013)659605.

\bibitem{47} Pandya D.M., Thomas V.O. and Sharma R.: Astrophys.
 Space Sci. \textbf{356}(2015)285.

\bibitem{50} Molina, A., Dadhich, N. and Khugaev, A.: Gen. Relativ. Gravit.
\textbf{49}(2017)1.

\bibitem{52} Banerjee, A., Jasim, M.K. and Pradhan, A.: Mod. Phys.
Lett. A \textbf{35}(2020)2050071.

\bibitem{f2bb} Dayanandan, B. et al.: Chin. J. Phys. \textbf{82}(2023)155.

\bibitem{f2aa} Gul, M.Z. et al.: Eur. Phys. J. C \textbf{84}(2024)8.

\bibitem{3gg} Mustafa, G. et al.: Chin. J. Phys. \textbf{88}(2024)954.

\bibitem{f61} Rej, P., Bogadi, R.S. and Govender, M.: Chin. J. Phys.
\textbf{87}(2024)608.

\bibitem{f62} Shahzad, M.R. et al.: Phys. Dark Universe \textbf{46}(2024)101646.

\bibitem{f64} Das, B. et al.: Astrophys. Space Sci. \textbf{369}(2024)76.

\bibitem{53a} Weyl, H.: Sitzungsber. Preuss. Akad. Wiss.
\textbf{465}(1918)01; Dirac, P.A.M.: Proc. R. Soc. London A
\textbf{333}(1973)403.

\bibitem{1ab} Schertler, K. et al.: Nucl. Phys. A \textbf{677}(2000)463;
Yan, Y. et al.: Phys. Rev. D \textbf{86}(2012)114028.

\bibitem{aa} Mandal, S., Sahoo, P.K. and Santos, J.R.: Phys. Rev. D
\textbf{102}(2020)024057.

\bibitem{bb} Khyllep, W., Paliathanasis, A. and Dutta, J.: Phys. Rev. D
\textbf{103}(2021)103521.

\bibitem{cc} Lin, R.H. and Zhai, X.H.: Phys. Rev. D \textbf{103}(2021)124001.

\bibitem{dd} Zhao, D.: Eur. Phys. J. C. \textbf{82}(2022)303.

\bibitem{55} Sharma, R. and Maharaj, S.D.: Mon. Not. R. Astron. Soc.
\textbf{375}(2007)1265; Ngubelanga, S.A., Maharaj, S.D. and Ray, S.:
Astrophys. Space Sci. \textbf{357}(2015)1; Abbas, G. and Nazar, H.:
Ann. Phys. \textbf{424}(2021)168336.

\bibitem{56} Cheng, K.S., Dai, Z.G. and Lu, T.: Int. J. Mod. Phys. D
\textbf{7}(1998)139.

\bibitem{57} Chodos, A.: Phys. Rev. D \textbf{9}(1974)3471.

\bibitem{58} Witten, E.: Phys. Rev. D \textbf{30}(1984)272.

\bibitem{59} Farhi, E. and Jaffe, R.L.: Phys. Rev. D \textbf{30}(1984)2379.

\bibitem{59a} Peshier, A., Kampfer, B. and Soff, B.: Phys. Rev. C
\textbf{61}(2000)045203.

\bibitem{59b} Rehberg, P., Klevansky, S.P. and H$\ddot{u}$fner, J.: Phys. Rev. C
 \textbf{53}(1996)410. Hanauske, M. et al.: Phys. Rev. D \textbf{64}(2001)043005;
 R$\ddot{u}$ster, S.B. and Rischke, D.H.: Phys. Rev. D \textbf{69}(2004)045011;
 Menezes, D.P., Provid$\hat{e}$ncia, C. and Melrose, D.B.: J. Phys. G: Nucl. Part. Phys.
 \textbf{32}(2006)1081; Jiang, Y. et al.: Phys. Rev. D
 \textbf{85}(2012)034031.

\bibitem{59c} Baluni, V.: Phys. Rev. D \textbf{17}(1978)2092; Fraga, E.S., Pisarski,
R.D. and Schaffner-Bielich, J.: Phys. Rev. D
\textbf{63}(2001)121702.

\bibitem{As} Aghanim, N. et al.: Astron. Astrophys. \textbf{652}(2021)C4.

\bibitem{60} Deb, D. et al.: Mon. Not. R. Astron. Soc. \textbf{485}(2019)5652;
Rahaman, F. et al.: Phys. Rev. D \textbf{82}(2010)104055; Kalam, M.
et al. : Eur. Phys. J. C \textbf{72}(2012)2248; Rahaman, F. et al.:
Gen. Relativ. Gravit. \textbf{44}(2012)107.

\bibitem{61} O'Brien, S. and Synge, J.L.: Commun. Dubl. Inst. Adv. Stud.
\textbf{9}(1952)1.

\bibitem{62} $\ddot{O}$zel, F., G$\ddot{u}$ver, T. and Psaltis, D.: Astrophys. J.
\textbf{693}(2009)1775.

\bibitem{63} Rawls, M.L. et al.: Astrophys. J. \textbf{730}(2011)25.

\bibitem{es} Hell, T. and Weise, W.: Phys. Rev. C 90(2014)045801.

\bibitem{1a3} Buchdahl, A.H.: Phys. Rev. D \textbf{116}(1959)1027.

\bibitem{hy} Barraco, D.E. and Hamity, V.H.: Phys. Rev. D
\textbf{65}(2002)124028.

\bibitem{64} Gadbail, G.N., Mandal, S. and Sahoo, P.K.: Physics \textbf{4}(2022)1403.

\bibitem{65} Abreu, H., Hern$\acute{a}$ndez, H. and N$\acute{a}$nez, L.A.:
 Class. Quantum Grav. \textbf{24}(2007)4631.

\bibitem{66} Chandrasekhar, S.: Astrophys. J. \textbf{140}(1964)417;
Chan, R., Herrera, L. and Santos, N.: Mon. Not. R. Astron. Soc.
\textbf{265}(1993)533.

\bibitem{50a} Tolman, R.C.: Phys. Rev. \textbf{35}(1930)896;
Tolman, R.C.: Phys. Rev. \textbf{55}(1939)364; Oppenheimer, J.R. and
Volkoff, G.M.: Phys. Rev. \textbf{55}(1939)374.

\bibitem{66a} Bhar, P., Malik, A. and Almas, A.: Chin. J.
 Phys. \textbf{88}(2024)839; Bhar, P. and Pretel, J.M.: Phys. Dark Universe
 \textbf{42}(2023)101322; Kaur, S. et al.: New Astron. \textbf{110}(2024)102230.

\bibitem{67} Bhar, P. et al.: Phys. Dark Universe \textbf{46}(2024)101686.

\bibitem{68} Nazar, H. and  Abbas, G.:  Adv. Astron. \textbf{2021}(2021)6698208.

\bibitem{54} Sharif, M. and Ibrar, I.: Chin. J. Phys.
\textbf{89}(2024)1578; Eur. Phys. J. Plus \textbf{139}(2024)1; Phys.
Scr. \textbf{99}(2024)105034.

\end{thebibliography}
\end{document}